\documentclass[aps,prb,reprint,groupedaddress,amsmath,amssymb,preprintnumbers,showpacs,longbibliography]{revtex4-2}
\usepackage{graphicx}
\usepackage{hyperref}
\usepackage{amsmath,bm}
\usepackage{dcolumn}
\usepackage{braket}
\usepackage{physics}
\usepackage{float}
\usepackage{xcolor}
\usepackage[normalem]{ulem}
\usepackage{soul}

\newcommand{\be}{\begin{equation}}
	\newcommand{\ee}{\end{equation}}
\newcommand{\ba}{\begin{eqnarray}}
	\newcommand{\ea}{\end{eqnarray}}
\newcommand{\bs}{\begin{subequations}}
	\newcommand{\es}{\end{subequations}}
\newcommand{\bw}{\begin{widetext}}
	\newcommand{\ew}{\end{widetext}}

\usepackage[final]{showlabels}
\usepackage{cleveref}
\crefname{equation}{Eq.}{Eqs.}
\crefname{figure}{Fig.}{Figs.}
\Crefname{Figure}{Fig.}{Figs.}
\crefname{table}{Table}{Tables}
\crefname{section}{Sec.}{Secs.}
\crefname{appendix}{Appendix}{Appendix}
\Crefname{section}{Sections}{Sections}
\crefname{subsection}{Sec.}{Sec.}
\Crefname{subsection}{Subsection}{Subsections}
\crefname{subsubsection}{subsection}{subsections}
\Crefname{subsubsection}{Subsection}{Subsections}
\crefname{paragraph}{subsection}{subsections}
\Crefname{paragraph}{Subsection}{Subsections}

\begin{document}
	\title{Effects of lattice imperfections on high-harmonic generation from correlated systems}
	\date{\today}
	\author{Thomas Hansen and Lars Bojer Madsen}
	\affiliation{Department of Physics and Astronomy, Aarhus
		University, DK-8000 Aarhus C, Denmark}
	\begin{abstract}
		Using the one-dimensional Fermi-Hubbard model we study the effects of lattice imperfections on high-harmonic generation (HHG) from correlated systems. We simulate such imperfections by randomly modifying the chemical potential across the individual lattice sites. We control the degree of correlation by varying the Hubbard $U$. In the limit of vanishing $U$, this approach results in Anderson localization. For non-vanishing $U$, we explain the spectral observations using a qualitative picture in which correlation and the imperfections may balance each other out, causing Bloch-like transitions, i.e., transitions similar to those occurring in the case with small or vanishing $U$ and with vanishing imperfection-induced energy gaps, even though the dynamics take place under conditions of high $U$ and severe imperfections. We verify this picture by studying HHG spectra where imperfections are found to cause gain in the HHG spectra. The spectral gain is mainly in high harmonic orders for low correlation and low harmonic orders for high correlation. We explain this using the following qualitative picture. For low correlation, the addition of imperfections gives rise to significant energy gaps in the system, corresponding to high harmonics, and when the correlation is massive the addition of imperfections balances out the correlation, resulting in smaller energy gaps, meaning lower harmonic orders.
		We further demonstrate how in the unbalanced cases, i.e., high $U$ and relatively small amount of imperfection or vice versa, the dynamics are largely adiabatic, and how in the balanced cases, with similar magnitude of correlation and imperfection, the dynamics are qualitatively similar. 
	\end{abstract}
	\maketitle
	\section{Introduction}
	Studying the dynamics of solids on the ultrafast ($10^{-18}$s - $10^{-15}$s) time-scale is an immensely interesting possibility. However, the methods needed to achieve such a time resolution still need some development. One of the approaches is high-order harmonic generation (HHG). HHG is a highly nonlinear ultrafast process that may be used as a spectrographic tool with time resolution on the subfemtosecond timescale \cite{Li2008, Lein2002, Torres2007, Kraus2015, Luu2018, Silva2018}. HHG is also used as a method to produce ultrashort pulses with high photon energy \cite{Schubert2014, Lewenstein1994, Lein2003, Li2008, Ghimire2011}. The high particle density of solids may aid in the production of higher intensity pulses and thereby improve its usefulness for time-resolved measurements \cite{Schubert2014, Luu2018, Garg2016, Luu2015, You2017, Kaneshima2018, Liu2017, JensenPRA2022, Yamada2021, Hansen2017, Tancogne-Dejean2017, Floss2018, Murakami2021}. Describing HHG from solid bandgap materials is often done within the theoretical framework of bandstructure theory. Here HHG has been, qualitatively, separated into two categories, intra- and interband HHG. Intraband HHG is the generation of harmonics by the acceleration of charge carriers in a given band of the solid. Interband generation is perhaps best described using the so-called 3-step model: (1) an electron is excited across the bandgap between the bands in question, (2) the electron and generated hole propagate through their respective bands, this step results in intraband HHG, (3) the electron and hole recombine under the emission of a photon with energy corresponding to the energy difference between the bands at the crystal momentum of recombination. That photon is an interband high-harmonic \cite{Vampa2014, Golde2008}. 
	
	The band picture, used for the 3-step model, requires the translational symmetry gained by applying the mean-field approximation. The mean-field approximation allows the electrons to be treated as independent particles, which is not exact. With that in mind, over the last 5 years, researchers have worked on the effects of beyond mean-field electron-electron correlation on HHG both theoretically  \cite{Silva2018, Murakami2018, Murakami2021, Lysne2020, Tancogne-Dejean2018, Imai2020, Chinzei2020, Orthodoxou2021, Murakami2021, Shao2022, Hansen22, Masur2022, Udono22, Murakami2022, Hansen22_2} and experimentally \cite{Uchida2022, Granas2022, Bionta2021}. Much theoretical research \cite{Silva2018, Murakami2018_2, Murakami2021, Hansen22, Hansen22_2, Lysne2020} has used the Hubbard model, which is also used in this work. The Hubbard model includes electron-electron correlation through the Hubbard $U$, which energetically penalizes the occurrence of doubly-occupied sites. 
	Studies using the Hubbard model have led to the development of a parallel to the 3-step model but in the quasi-particle picture of doublons and holons \cite{Murakami2021}. Doublons are doubly occupied lattice sites and holons are empty lattice sites. Mott insulators, i.e., solids that are dominated by electron-electron correlation and with an uneven number of valence electrons per lattice site, are typically well described by this quasi-particle picture in the high-$U$ limit. The doublon-holon 3-step model parallel consists of the following steps, (1) a doublon-holon pair is created, (2) the doublon and holon propagate through the solid, and (3) they recombine \cite{Murakami2021}. The energetic penalty introduced through the $U$ term causes the creation of a doublon in the doublon-holon pair to require a significant amount of energy, equivalent to the excitation of an electron across a band gap. Similarly, the recombination of a doublon and holon corresponds to the recombination of the electron and hole from the original 3-step model. When the electron-electron correlation dominates compared to the hopping term, $t_0$, i.e., when $U/t_0\gg 0$, the eigenenergies of the Hubbard model arrange themselves into so-called Hubbard sub-bands, with the Mott gap being equivalent to the band gap. 
	
	The Hubbard model describes a perfectly translationally symmetric lattice, which is an idealization as lattice vibrations cause aperiodicity in the lattice. Moreover, defects and impurities would break the translational symmetry. Therefore, in this work, we introduce an extra term to the Hubbard Hamiltonian which captures the effects of such imperfections. The phenomenon known as Anderson localization arises from such lattice imperfections, and this effect significantly influences the conductivity of the system \cite{Anderson1958}. Physical phenomena and conditions that affect the conductivity affect the current and, therefore, the HHG spectrum. In the $U=0$ limit, HHG from lattices with lattice imperfections has been studied before \cite{Almalki2018, Orlando18, Chuan19ImperfectCrystals,Pattanayak2020,Mrudul2020,Iravani2020,pattanayak2021,Zeng2022}. In those studies, it was shown that imperfections, and the Anderson localization which results from them, may result in high-harmonic gain \cite{pattanayak2021}. The study of impurities has led to a 3-step model qualitatively similar to that of atomic/molecular HHG. In the first step, an electron or hole is excited away from the impurity it was bound to in the initial state, in the second step the now unbound electron or hole propagates within the lattice, and in the third step, it recombines with the impurity resulting in the emission of a high harmonic \cite{Almalki2018}. Now, the correlation breaks the translational symmetry of each electron, on which the Bloch picture is based, but the laser-driven Hubbard Hamiltonian still has a shift symmetry, i.e., a symmetry under shifting all electrons in the lattice by an integer number of lattice spacings. Lattice imperfections break this symmetry. As the ground state of the Hubbard Hamiltonian belongs to a specific eigenvalue of the shift operator, and all works that we know of have started from the Hubbard Hamiltonian's ground state, breaking this symmetry will also effectively increase the size of the Hilbert space substantially, which again could have interesting effects. Finally, the Hubbard model does facilitate some degree of many-body localization \cite{Prelov2016}, and so we do expect effects induced by breaking the shift symmetry, which could be indications of many-body localization. See Refs.~\cite{Alet2018, Abanin2019} for recent reviews on many-body localization.

	In this work we set out to answer the following questions: (i) How is the HHG spectra affected by the addition of lattice imperfections in the presence of correlation? (ii) How do the effects of lattice imperfections and electron-electron correlation interact? and (iii) How do the dynamics in the lattice relate to the spectra?
	
	The paper is organized as follows. In \cref{sec:Theoretical model and methods}, we introduce the theoretical model, the theoretical measures used, numerical methods, and introduce a qualitative picture of the system, which will be used throughout. In \cref{sec:Results}, we discuss and analyze our results. In \cref{sec:Conclusion}, we summarize and conclude.
	We use atomic units unless we explicitly state otherwise.
	\section{Theoretical model and methods} \label{sec:Theoretical model and methods}
	This section is separated into four subsections. Firstly, in \cref{sec:Hamiltonian}, we introduce the system Hamiltonian. Then, in \cref{sec:Qualitative physical picture}, we introduce a qualitative picture of the system, which will be utilized in the discussion. In \cref{sec:symmetries}, a shift symmetry of the system is discussed, and in \cref{sec:measures} we introduce the measures we use in the discussion. Finally, in \cref{sec:Simulation method}, we briefly describe the simulation method used and provide the system parameters.
	\subsection{Hamiltonian}\label{sec:Hamiltonian}
	In this work, we use the Hubbard model to simulate a one-dimensional $L=12$ atoms long chain of atoms with periodic boundary conditions at half filling, i.e., the number of electrons equals $L$. To study the effect of lattice imperfections on the process of HHG we add a term to the Hubbard Hamiltonian which treats the lattice imperfections and a modification to the Hopping term which treats the laser field from which the high harmonics are generated. Treating the incoming laser pulse in the dipole approximation, i.e., neglecting its special dependence results in the following Hamiltonian
	\begin{align}
		\hat{H}(t)&=\hat{H}_\text{Hub}(t)+\hat{H}_\Delta, \label{eq: Hamiltonian}
	\end{align}
	with the laser-driven Hubbard Hamiltonian given by
	\begin{align}
		\hat{H}_\text{Hub}(t)&=	\hat{H}_\text{Hop}(t)+\hat{H}_U, \label{eq: Hubbard Hamiltonian}
	\end{align}
	where
	\begin{align}
		\hat{H}_\text{Hop}(t)&=-t_0\sum_{i,\sigma}\left(e^{iaA(t)}\hat{c}_{i+1,\sigma}^\dagger \hat{c}_{i,\sigma}+\mathrm{h.c.}\right)\\
		\hat{H}_U&=U\sum_{i}\hat{n}_{i,\uparrow}\hat{n}_{i,\downarrow}.
	\end{align}
	In Eq.~(\ref{eq: Hamiltonian}),
	\begin{align}
		\hat{H}_\Delta&=\sum_{i}\Delta A_i\left(\hat{n}_{i,\uparrow}+\hat{n}_{i,\downarrow}\right), \label{eq:H Delta}\\
		A_i&=\cos(\phi_i) \label{eq:Ai}.
	\end{align}
	Here $\hat{H}_\text{Hop}$ is the hopping part of the Hamiltonian, $\hat{H}_U$ is the correlation-part of the Hamiltonian, and $\hat{H}_\Delta$ is the part which treats the lattice imperfections. Within $\hat{H}_\text{Hop}$, $-t_0$ is the transition matrix element corresponding to nearest neighbor hopping, $e^{iaA(t)}$ is Peierl's phase, and $c^\dagger_{i,\sigma}$ and $c_{i,\sigma}$ are the operators for creation and annihilation of an electron on site $i$ with spin $\sigma\in\{\uparrow,\downarrow\}$, respectively. In Peierl's phase, $a$ is the lattice spacing and $A(t)$ is the vector potential of the incoming laser field in the dipole approximation. In $\hat{H}_U$, $U$ is the beyond-mean-field energy correction to the mean-field potential which results from a doubly occupied site, i.e., a doublon, and $\hat{n}_{i,\sigma}=\hat{c}^\dagger_{i,\sigma}\hat{c}_{i,\sigma}$ is the electron counting operator, which counts the electron on site $i$ with spin $\sigma$. Finally in $\hat{H}_\Delta$, $\Delta A_i$ is the energy needed to add an electron to site $i$. We pick the $A_i$ values according to \cref{eq:Ai} by evaluating cosine of a random, uniformly distributed phase $\phi_i\in\left[0;2\pi\right]$, and finally in \cref{eq:H Delta} scale $A_i$ by the system parameter $\Delta$. The energy axis has been picked such that $\langle A_i\rangle=0$. All simulations are generated using the same set of $\phi_i$-values unless it is explicitly stated otherwise. We treat $\Delta$ in the same manner we do with $U$, that is as a parameter.
	
	Ordinarily in the perfect lattice the onsite matrix element is set to zero, by picking the zero-point energy accordingly \cite{Silva2018, Murakami2021}. This effectively sets the chemical potential to zero which by the symmetry is site independent. A way to describe the effect of $\hat{H}_\Delta$ in physical terms is that the chemical potential at each site is modified by $\Delta A_i$, see Fig.~\ref{fig:illustration}. 
	
	The laser pulse is taken as an $N_c=10$ cycle laser pulse with a $\sin^2$ envelope. The laser vector potential is expressed as 
	\begin{align}
		A(t)&=A_0\cos(\omega_Lt-N_c\pi)\sin^2\left(\frac{\omega_Lt}{2N_c}\right). \label{eq:A}
	\end{align}
	Here $A_0$ is the vector-field laser amplitude, and $\omega_L$ is the laser carrier frequency.

	\subsection{Qualitative physical picture} \label{sec:Qualitative physical picture}
\begin{figure}
			\includegraphics[width=\linewidth]{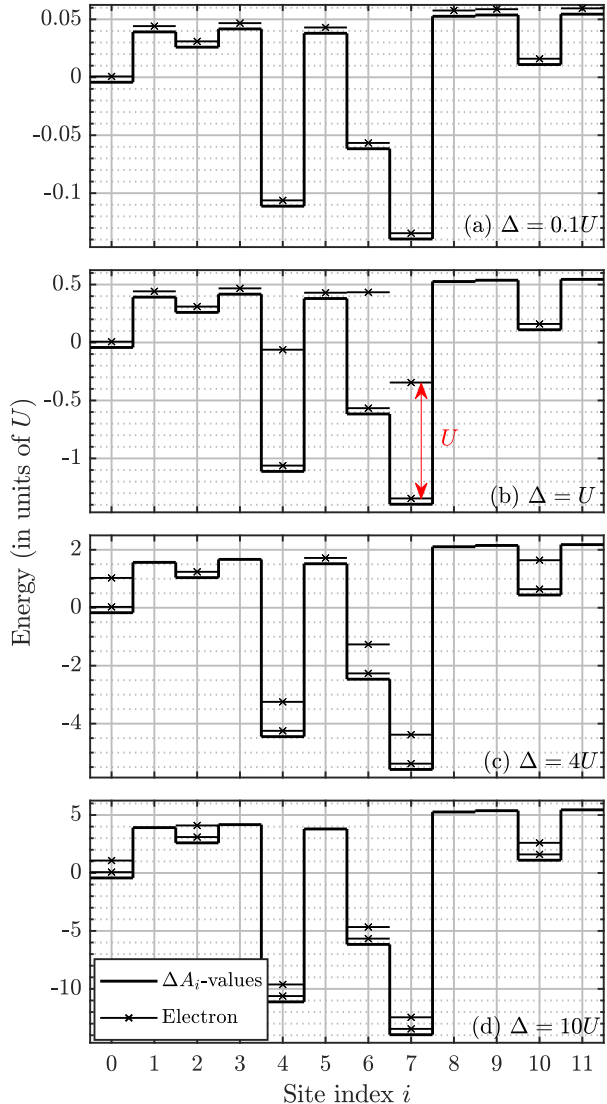}
			\caption{Illustration of the dominant ground state configurations of systems in the limit of $t_0\ll (U,\Delta)$. The thick black lines represent the modifications to the chemical potential at the various sites, and the thin, short black lines with crosses represent electrons. The second electron placed on a given site is displaced upwards by $U$, in order to represent the energy cost of adding a doublon, see the double-headed arrow in panel (b).  }
			\label{fig:illustration}
		\end{figure}

	In this section, we introduce a physical framework in which we can discuss the system and its dynamics. In the limit of $t_0\ll \max(U,\Delta)$ the eigenstates of the systems converge toward individual electron configurations. As the configurations are the basis states used, it is trivial to find the configuration associated with the ground state. Analysis of those configurations gives insight into the possible dynamics of the system. We have illustrated such dominant ground state configurations in \cref{fig:illustration} for four different values of $\Delta$ in terms of $U$ for a specific realization of the set of $\left\{A_i\right\}$ values in Eq.~(\ref{eq:Ai}). Whether one thinks of the change of $\Delta$ in terms of $U$, as $\Delta$ changing or as $U$ changing of course gives exactly the same figure, and results in the same conclusions. Notice the significantly varying ordinate scale in the panels. In the limit of $U\gg (\Delta,t_0)$ this system is a Mott insulator, due to the half-filling of the system.
	
	For $\Delta=0.1U$, \cref{fig:illustration} (a), the effect of $\hat{H}_\Delta$ is almost insignificant, notice the scale of ordinate. As a result $\hat{H}_U$ dominates the system, resulting in each site being populated equally, which, for the half-filling case used here, means each site is populated by one electron. Generally, the electrons are arranged in antiferromagnetic ordering, i.e., alternating between spin up and down on each site, in the correlation-dominated limit. This ordering means any transition across nearest-neighbor lattice sites will create a doublon-holon pair, which requires an energy of around $U$. For sufficiently large $U$, compared to $\Delta$, that energy requirement can effectively lock all the electrons resulting in very little dynamics and therefore only little harmonic generation \cite{Silva2018, Hansen22_2, Murakami2018_2, Murakami2021}. In this limit, the doublon-holon based 3-step model captures the physics as $\Delta$ has very little effect. For $\Delta=U$, \cref{fig:illustration} (b), $\hat{H}_\Delta$ is significant enough to create three doublons, on sites 4, 6, and 7, in spite of the $U$-term which penalizes the creation of doublons energetically. This makes it significantly easier to cause the electron on site 10 to hop to either site 9 or 11 than it was in the $\Delta=0.1U$ case. Another way to view this effect is in terms of the 3-step model for impurities. The addition of $U$ to a system with significant $\Delta$ effectively lowers the binding energy of the electrons to the imperfection likely resulting in harmonic gain but at lower harmonic orders due to the lowered binding energy. For higher $\Delta$-values more and more doublons are created in the ground state, as seen in \cref{fig:illustration} (c), where $\Delta=4U$, and doublons have been created on site 0, 4, 6, 7, and 10, and finally in \cref{fig:illustration} (d), where $\Delta=10U$, and all electrons are in doublons, on the sites, 0, 2, 4, 6, 7, and 10. This trend illustrates the interplay between $\hat{H}_\Delta$, which pushes the electrons towards the sites with lowest $A_i$-values and $\hat{H}_U$, which energetically penalizes the presence of doublons, and therefore pushes the electrons to distribute themselves with one electron per site. Describing this in terms of the impurity-based 3-step model, the increasing $\Delta$ corresponds to the lowered binding energy which could increase the maximal harmonic orders created but at the same time lower the gain from the given imperfection.
		
	It is also worth paying attention to the transition from site 7 to 8. In \cref{fig:illustration} (d), $\Delta=10U$, exciting an electron from site 7 to 8 would require $\simeq18.2U=1.82 \Delta$ of energy, whereas for the smaller $\Delta=4U$, seen in \cref{fig:illustration} (c), it only requires $6.68U=1.67 \Delta$ and for $\Delta=U$, seen in \cref{fig:illustration} (b), the gap is down to $0.92U=0.92 \Delta$, finally for $\Delta=0.1U$ requires $1.19U=11.9\Delta$. It is therefore possible that increasing $\Delta$ in terms of $U$ will lower the energy required for certain transitions. When $\Delta\gg U$ the binding energy is massive, it then falls as $U$ increases, until the $U$ term results in a more even electron distribution across the lattice meaning the doublon-holon-based 3-step model becomes the appropriate picture. 
	
	In conclusion, it is possible, as illustrated in \cref{fig:illustration}, for the electron-electron repulsion of $\hat{H}_U$ to partially balance out the attraction to certain sites induced by $\hat{H}_\Delta$ and thereby make it easier to drive electrons across sites. This mechanism works even for large $U$, however, $\Delta$ needs to be large enough to create the holons needed to facilitate transitions that are not energetically penalized by $\hat{H}_U$. This can be seen, e.g., between site 9 and 10 in \cref{fig:illustration} (c), where site 9 is empty, due to $\Delta$'s size, this means one of the electrons in the doublon on site 10 has approximately the same energy it would have if it sat on site 9 instead. It is therefore easy for the pulse to drive an electron from site 10 to site 9 in that case.  If $\Delta$ is noticeably smaller than $U$, then every site would have precisely one electron, and every transition would require the creation of a doublon, which has an $\hat{H}_U$-associated energy-cost of $U$.  
	
	Finally, we point out the dynamics as those associated with site 5 in Figs.~\ref{fig:illustration} (b) and (c). The addition of doublons to the ground state increases the amount of Pauli blocking in the system, which causes the electron on site 5 to be blocked, it cannot move anywhere due to the doublons on sites 4 and 6 and the nearest-neighbor hopping approximation. 
	
	While the above discussion strictly speaking only pertains to the limit where $\hat{H}_\text{Hop}$ is insignificant, the general understanding of $\hat{H}_\Delta$ pushing the electrons to occupy certain sites, and $\hat{H}_U$ pushing the electrons away from each other, typically resulting in antiferromagnetic ordering, applies regardless of the value of $t_0$ relative to the values of $U$ and $\Delta$. 
		
	\subsection{Symmetries}\label{sec:symmetries}
	
	The Hubbard Hamiltonian, $\hat{H}_\text{Hub}(t)$ of Eq.~(\ref{eq: Hubbard Hamiltonian}), has an important symmetry, which is broken by the addition of $H_\Delta$. When periodic boundary conditions are applied to the field-free or laser-driven Hubbard Hamiltonian it is symmetric under the operation which shifts all electrons by one lattice spacing, i.e., takes the site index $i$ to $i+1$ or vice versa \cite{Minahan2006, Hubbard}. The periodic boundary conditions imply that site $L$ is equivalent to site $0$, enumerating the sites from 0 to $L-1$. Defining the operator which takes $i$ to $i+1$ as $\hat{S}$, we have $\left[\hat{H}_\text{Hub}(t),\hat{S}\right]=0$, which means all eigenstates can be picked such that they are eigenstates of both $\hat{H}_\text{Hub}(t)$ and $\hat{S}$ simultaneously. Letting $\ket{\psi_i}$
	be such an eigenstate implies $\hat{S}\ket{\psi_i}=s_i\ket{\psi_i}$. The periodic boundary condition further implies that $\hat{S}^L=\hat{I}$, where $\hat{I}$ is the identity operator. This relation further implies that $s_i^L=1$, which means all $s_i$'s can be expressed as $\exp(iK_ia)$. Expressing $s_i$ in this manner means that $s_i^L=1$ can be rewritten in the form $LaK_i=2\pi i\Leftrightarrow K_i=2\pi i/La$, which are the same $k$-values as those observed in Bloch phases arising from Bloch's theorem. The only difference to the Bloch case is that the Bloch states arise from symmetry under translation of any individual electron, whereas in the present case, the $K_i$'s arise from the shift, or translation, of all electrons together. This difference also implies that $K_i$ is the total crystal momentum in the given state.  We remark that by the nature the exponential function $2\pi n/La$ is completely equivalent to $2\pi (L+n)/La$ so that here are $L$ distinct values of $n$, e.g., $n\in\left\{0,1,2,\dots ,L-1\right\}$. 
	It turns out that in our realization, the Hubbard Hamiltonian for the 12-site long one-dimensional lattice, has a degenerate ground state which includes states with $K_i=0$ and $K_i=6\times2\pi/La=\pi/a$, where the last equality applies to our $L=12$ system. These values of $K_i$ are completely determined by $\hat{H}_\text{Hop}$, as $\hat{H}_U$ represents total crystal-momentum conserving scattering events, which do not change the total crystal momentum, $K_i$. We note that this shift symmetry is not dependent on the nearest-neighbor approximation, the one-band approximation taken in $\hat{H}_\text{Hop}$, nor the onsite approximation taken in $\hat{H}_U$. However, the $\hat{H}_\Delta$ term does break the shift symmetry by construction. 
	
	\subsubsection{Construction of eigenbasis for shift operator}
	
	In the discussion in \cref{sec:Results}, the population in states with specific $K$ values are plotted and discussed. In order to do that an eigenbasis of the shift operator is needed. Such a basis can be created by first taking an arbitrary electron configuration, shifting it one lattice spacing at a time, till one ends up in the original configuration. Then one creates linear combinations of the configurations one reached by shifting, multiplying each term in the sum with the appropriate phase, $\exp(inaK_i)$, where $n$ is the number of sites the configuration has been shifted by. Doing this for all configurations, which cannot be shifted into one another, results in a complete basis of the system with each basis vector being an eigenstate of the shift operator, with eigenvalue given by the $K_i$ value chosen in the phase. 
	\subsection{Measures}\label{sec:measures}

	In this work, we consider a system that is thin in the direction of the laser propagation. This allows us to neglect propagation effects \cite{Gaarde2008}. When propagation effects can be neglected the spectrum becomes proportional to the norm-square of the electron acceleration \cite{Baggesen2011}. This proportionality means we can model the spectrum, $S(\omega)$, as
		\begin{align}
			S(\omega)=\left|\mathcal{F}\left(\frac{d}{dt}j(t)\right)\right|^2=\left|\omega j(\omega)\right|^2, \label{eq:Spetrum}
		\end{align}
	where $j(\omega)$ is the Fourier transform, $\mathcal{F}$, of the current $j(t)$. The current can be determined via Heisenberg's equation of motion \cite{Mahan}, resulting in
	\begin{align}
		j(t)=\left\langle-iat_0\sum_{i,\sigma}\left(e^{iaA(t)}\hat{c}_{i+1,\sigma}^\dagger \hat{c}_{i,\sigma}-\mathrm{h.c.}\right)\right\rangle. \label{eq:j}
	\end{align}

	As $\hat{H}_\Delta$ breaks the translational symmetry of the system, we will analyze the population on each individual site of the lattice. To illustrate the most prudent points we introduce the following measure,
	\begin{align}
		C_\text{pop}^\Delta(t)=\frac{1}{L}\sum_{i=1}^L (P_i(t)-\left\langle P\right\rangle)A_i. \label{eq:C_pop^And}
	\end{align}
	This measure calculates the covariance between the electron population in the lattice, given by the $P_i$ values, and the $A_i$ values. Here $P_i(t)=\langle \hat{n}_{i,\uparrow}+\hat{n}_{i,\downarrow}\rangle$ is the time-dependent population on site $i$. Note that the $A_i$ values have been picked such that $\left\langle A_i\right\rangle=0$, and that $\left\langle P\right\rangle=N_\text{elec}/L$, where $N_\text{elec}$ is the number of electrons in the lattice. As the lattice is half-filled throughout this work, $N_\text{elec}=L\Leftrightarrow\left\langle P\right\rangle=1$. The measure in \cref{eq:C_pop^And} describes the degree to which the electrons are at the sites with minimal $A_i$-values. If those sites are the only ones populated then $C_\text{pop}^\Delta(t)$ reaches its minimal value, which is negative. If, on the other hand, every site is equally populated then $C_\text{pop}^\Delta(t)=0$. The eigenstates of the Hubbard Hamiltonian are equally populated across all sites when periodic boundary conditions are applied. 
	This means that only $\hat{H}_\Delta$ can cause $C_\text{pop}^\Delta(t)$ to become less than zero.
	Therefore, this measure provides a means to determine the importance of $\hat{H}_\Delta$, and through that a picture of how the electrons distribute themselves throughout the lattice. 
	
	We will also be using the following fidelity measure,
	\begin{align}
		W(t)=\left|\bra{\Psi(0)}\ket{\Psi(t)}\right|^2, \label{eq:W}
	\end{align}
	which is the norm square of the projection of the state at a given time $\ket{\Psi(t)}$ onto the initial state $\ket{\Psi(0)}$, which is the ground state of the field-free system. Calculating $W(t)$ gives us the opportunity to look at the speed with which (and the degree to which) the system leaves the ground state, which may give further insight into the nature of the dynamics in the systems.
	
	\subsection{Simulation method and parameters}\label{sec:Simulation method}
	We solve the time-dependent Schrödinger equation $\mathrm{i} \frac{d}{dt}\ket{\Psi(t)}=\hat{H}\ket{\Psi(t)}$ for $\ket{\Psi(t)}$ expressed in terms of configurations $\ket{\Phi_i}$, with $i$ being a collective index specifying the distribution of all the electrons on the sites, i.e., $\ket{\Psi(t)}=\sum_{i}c_i(t)\ket{\Phi_i}$.
	
	The simulations are done using the Arnoldi-Lancoz algorithm see, e.g., Refs.~\cite{Park1986, Smyth1998,Guan2007, Frapiccini2014}. We utilize imaginary time propagation to find the ground state, $\ket{\Psi(0)}$, which is used as the initial state of the simulations. The imaginary time propagation always starts from a $K=0$ state (see \cref{sec:symmetries}). This means the ground state will also have $K=0$, if $\Delta=0$, otherwise the ground state can, at least in principle, belong to any or all $K$ values. In order to test for convergence we calculate the energy and $D=\left\langle\sum_i \hat{n}_{i,\uparrow}\hat{n}_{i,\downarrow}\right\rangle/L$ and $\eta=\left\langle\sum_{i}\hat{S}_i\hat{S}_{i+1}\right\rangle/L$, with $\hat{S}_i=\left(\hat{n}_{i,\uparrow}-\hat{n}_{i,\downarrow}\right)/2$ being the spin operator for electrons on site $i$. These measures are used and discussed further in Ref.~\cite{Silva2018}. We only accept convergence when the change in the measures over 100 timesteps is limited by machine precision. The real-time propagation convergence is tested by comparing spectra and currents for different Krylov subspace dimensions and time steps. For convergence the maximal Krylov subspace dimension is 4 with a time step size of $1/\sqrt{10}$ a.u.
	
	The system parameters were chosen to describe Sr$_2$CuO$_3$ \cite{Tomita2001}, as was done previously in Ref.~\cite{Silva2018}. The specific values are $a=7.5589$ a.u. and $t_0=0.0191$ a.u. The laser frequency has been picked to $\omega_L=0.005$ a.u.$=33$ THz. The amplitude was picked to $A_0=F_0/\omega=0.0.194$ a.u., which corresponds to a peak intensity of $3.3\cross 10^{10}$ W/cm$^2$. The Hubbard $U$ and the disorder strength $\Delta$ are treated as parameters, as we mentioned in \cref{sec:Hamiltonian}.
	
	\section{results}\label{sec:Results}
	In this section, we show and discuss our results. This work is based on results obtained for a large scan of $U$ and $\Delta$ values. We utilize only a representative subset of the results to represent our findings.
	
	\subsection{$C_\text{pop}^\Delta$  - covariance results}
\begin{figure}	
		\includegraphics[width=\linewidth]{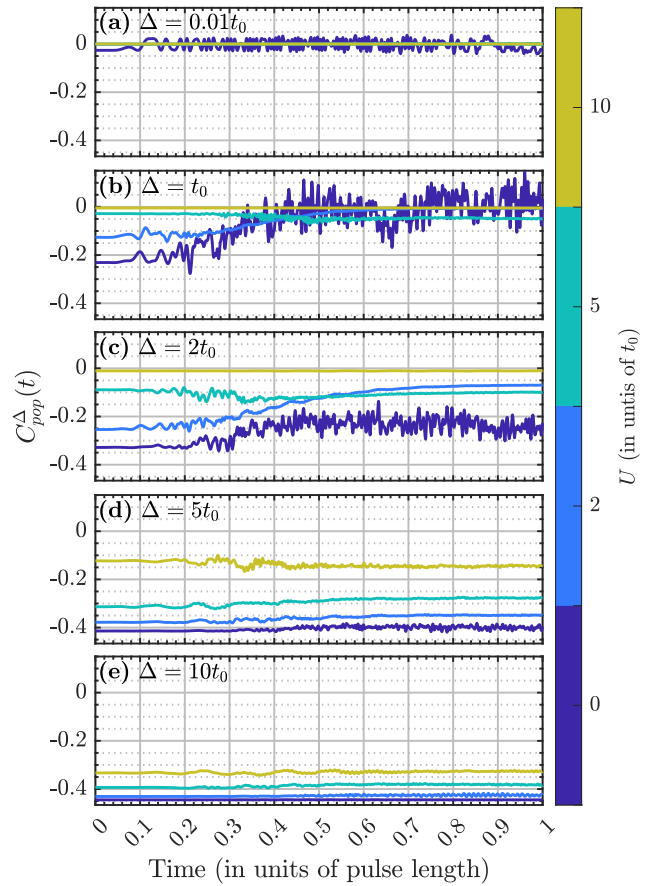}
		\caption{ $C_\text{pop}^\Delta(t)$ from \cref{eq:C_pop^And}. Each panel contains results for different $\Delta$ values, (a) $\Delta=0.01t_0$, (b) $\Delta=t_0$, (c) $\Delta=2t_0$, (d) $\Delta=5t_0$, and (d) $\Delta=10t_0$ for 4 different $U$ values, $0$, $2$, $5$, and  $10t_0$ as indicated by the colorbar.}
		\label{fig:C_pop^And}
	\end{figure}

	Figure~\ref{fig:C_pop^And} shows the covariance $C_\text{pop}^\Delta(t)$ measure for various $U$ and $\Delta$ values. Each panel contains results with the $U$ values $0$, $2$, $5$, and $10t_0$, while each panel has results with specific $\Delta$-values, (a) $\Delta=0.01t_0$, (b) $\Delta=t_0$, (c) $\Delta=2t_0$, (d) $\Delta=5t_0$, and (e) $\Delta=10t_0$. In \cref{fig:C_pop^And} (a), $\Delta=0.01t_0$, all  $C_\text{pop}^\Delta(t)$ values are very close to 0, some at times even slightly above it. The $U=0$ case shows slight variation, which makes sense as the lack of correlation means $\hat{H}_\Delta$ is the only thing that energetically favors some configurations over others. For the higher $U$ values, the results are all so close to each other that only the $U=10t_0$ result is visible. Since $\hat{H}_U$ energetically penalizes placing two electrons on one site, $\hat{H}_U$ pushes the system into $C_\text{pop}^\Delta(t)=0$ states. As $\Delta$ increases further in Figs.~\ref{fig:C_pop^And} (b), $\Delta=t_0$, and (c) $\Delta=2t_0$, the  $C_\text{pop}^\Delta(0)$ values decrease noticeably for all but the highest $U$-value, i.e., $U=10t_0$. In order for  $C_\text{pop}^\Delta(t)$ to decrease, the population on the sites with the lowest $A_i$-values needs to increase, i.e., the probability of observing a doublon on that site needs to increase. Doublons are energetically penalized via $\hat{H}_U$, which is dominant when $U=10t_0$. Hence in this case it is energetically favorable for such relatively small $\Delta$-values as used in Figs.~\ref{fig:C_pop^And} (b) and (c) for the electrons to distribute themselves evenly across the lattice. For the lower $U$-values this is not the case and some intermediate scenario between the completely even population distribution favored by $\hat{H}_U$, and the uneven distribution favored by $\hat{H}_\Delta$ is observed. As the pulse excites the system, we observe an increase in $C_\text{pop}^\Delta(t)$ resulting from the pulse driving the electrons away from the sites with the lowest $A_i$-values. Such electron movements result in a more even distribution of electrons across the lattice and therefore an increase in $C_\text{pop}^\Delta(t)$.
	
	As $\Delta$ increases further, to 5 and 10 $t_0$ in Figs.\ref{fig:C_pop^And} (d) and (e) respectively, the electrons become increasingly forced onto the sites with lowest $A_i$-values, resulting in a decreasing $C_\text{pop}^\Delta(t)$. In line with the picture discussed in \cref{sec:Qualitative physical picture}, the results for the lowest $U$-values become increasingly constant as $\Delta$
	increases see, e.g., the $U=0$ results, but the results for higher $U$ show the inverse scaling, i.e., they change more and more as $\Delta$ increases, see the $U=10t_0$ results. This comes as a result of $\hat{H}_\Delta$ and $\hat{H}_U$ balancing each other out when $U\approx \Delta$, which decreases the energy needed for some transitions. In this sense, the analysis of $C_\text{pop}^\Delta(t)$ supports the qualitative picture of \cref{sec:Qualitative physical picture}.
	
	\subsection{Spectra}
	\begin{figure}
		\includegraphics[width=\linewidth]{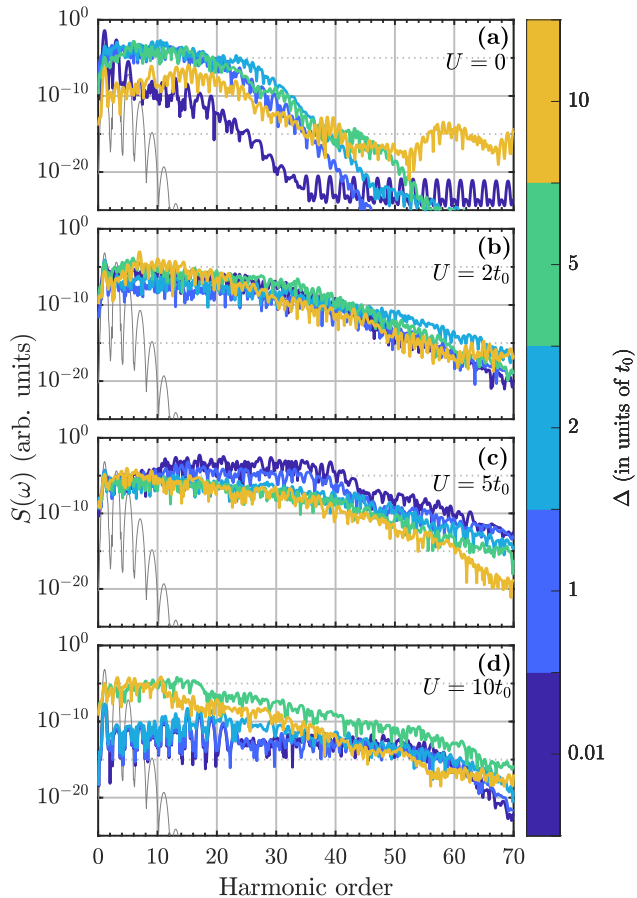}
		\caption{HHG spectra [$S(\omega)$ of \cref{eq:Spetrum}] for various values of $\Delta$ and $U$. (a) $U=0$, (b) $U=2t_0$, (c) $U=5t_0$, and (d) $U=10t_0$. The $U=\Delta=0$ spectrum, corresponding to that of intraband harmonics for Bloch electrons, is plotted in each panel as a thin grey line.}
		\label{fig:spectra}
	\end{figure}
		In \cref{fig:spectra}, HHG spectra for various $U$ and $\Delta$ values are shown. In \cref{fig:spectra} (a), $U=0$, the addition of $\hat{H}_\Delta$ has a rather significant impact on the spectra. It seems that similarly to the effects of adding correlation, the addition of lattice imperfections results in increased gain, particularly for low to medium harmonic orders, see \cref{fig:spectra} (a). This is likely a result of the prevalence of relatively weakly bound electrons, which generate a significant amount of lower harmonic orders. We note that the $U=\Delta=0$ case, as shown in the thin grey line, corresponds to intraband HHG, which explains the low gain, and the rapid spectral drop-off.
		In \cref{fig:spectra} (a), we also see that a large $\Delta$ value results in lowered gain in the lower harmonics, but continues to provide gain in the higher harmonics. This is as predicted in \cref{sec:Qualitative physical picture}: imperfections with large binding energies result in higher harmonics but less gain than imperfections with lower binding energies. As we randomize the potential on every site of the lattice, the lattice contains imperfections with reasonably low binding energies, even when $\Delta$ is large, see, e.g., site 2 in \cref{fig:illustration}.  A similar trend is also observed from $\Delta=0$ systems with increasing $U$ \cite{Silva2018,Murakami2021,Hansen22_2}. 
		In the limit $U\gg\Delta$ the appropriate physical picture is the doublon-holon-based 3-step model \cite{Murakami2021}. Essentially, when $U$ is large there is one electron on each site, meaning each one-electron transition creates a doublon which requires a significant amount of energy when $U$ is significant. For the $U=10t_0$ and $\Delta=0.01t_0$ system, it requires the equivalent of 26 harmonic orders which is a significant amount. In all cases with $\Delta \ne 0$, space-inversion symmetry is broken and even harmonics are observed.
		\begin{figure*}
			\includegraphics[width=.95\linewidth]{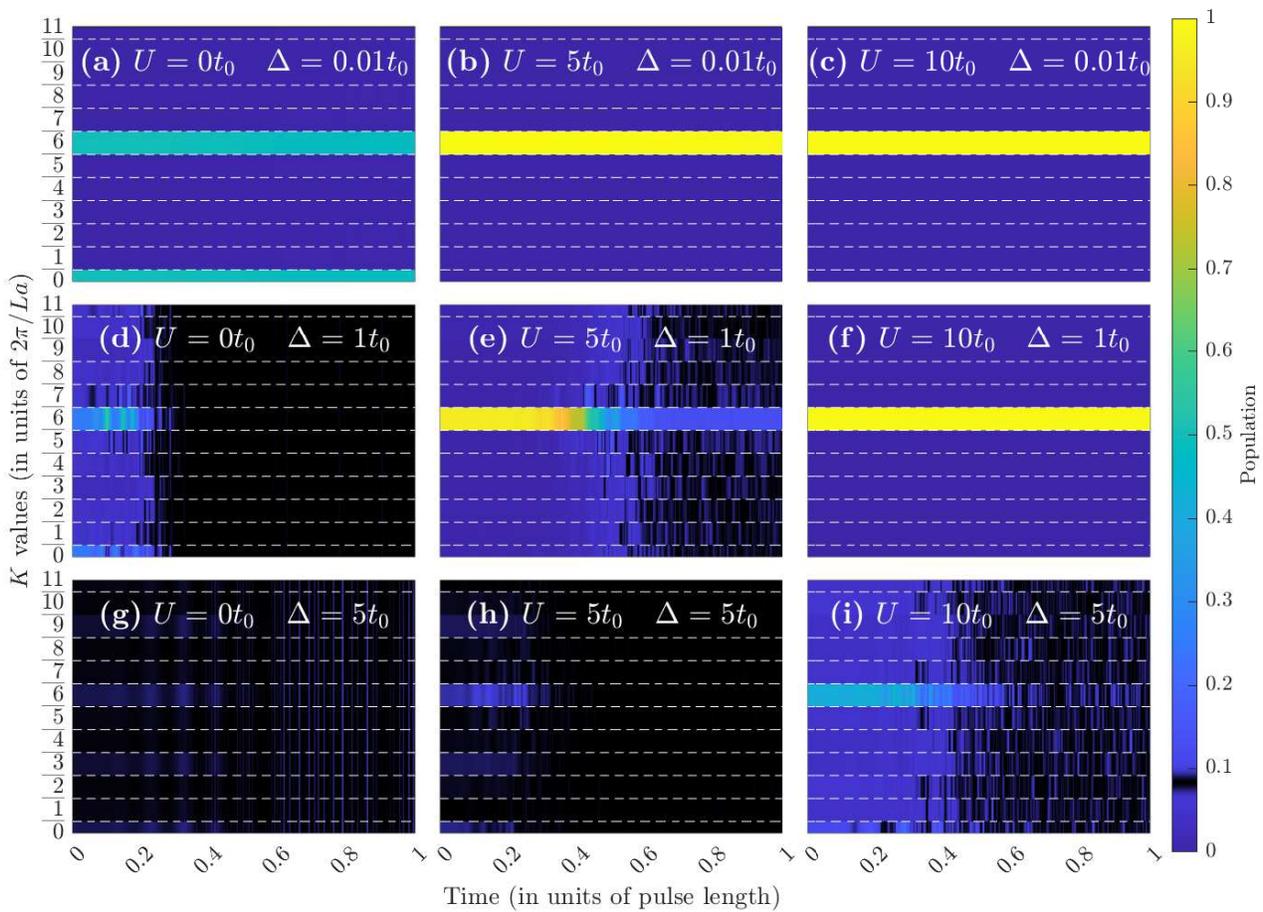}
			\caption{Population in configurations with different $K$ values over time. The panels show results for three $\Delta$ values, $\Delta=0.01t_0$ in (a), (b), and (c), $\Delta=t_0$ in (d), (e), and (f), and $\Delta=5t_0$ in (g), (h), and (i) and results for three different $U$ values, $U=0$ in (a), (d), and (g), $U=5t_0$ in (b), (e), and (h), and $U=10t_0$ in (c), (f), and (i). The colorbar has a black patch centered at $1/L$ which corresponds to equal distribution of the population across all $K$-values. This black color eases the identification of situations with equal population of all $K$-values.}
			\label{fig:rhos}
		\end{figure*}

		As $U$ increases, the electrons become bound in antiferromagnetic ordering. This means each electron is bound by the electrons around it rather than by the impurities, meaning the doublon-holon picture of HHG is a better description of the HHG process in this system. We note that the structure of the HHG spectra in the $\Delta=0$ limit has been explained thoroughly in terms of the doublon-holon pair picture in Refs.~\cite{Murakami2021, Hansen22_2}. If both $\Delta$ and $U$ are large, then it becomes possible to have a doublon situated at an imperfection in the ground state, see Fig.~\ref{fig:illustration}. Ordinarily, large $\Delta$ would mean large binding energy and therefore high harmonics being generated, but the large $U$ reduces that binding energy significantly resulting in the high harmonic gain for low harmonic orders with increasing $\Delta$ observed in \cref{fig:spectra} (d). In the intermediary $U$ values of Figs.~\ref{fig:spectra} (b), $U=2t_0$, and \ref{fig:spectra} (c), $U=5t_0$, the $U$ value is not so large as to decrease the system dynamics significantly, but large enough that it significantly impacts the binding of electrons to the lattice imperfections. In \cref{fig:spectra} (b) this manifests itself as very little change in the spectra regardless of the degree of impurity, as seen by the overlap of the spectra with various $\Delta$ values. This lack of change is likely a result of the amount of energy required to create a doublon being similar to the energy required to remove an electron from a binding imperfection. In \cref{fig:spectra} (c) the differences between the spectra with different $\Delta$ values are more pronounced than in \cref{fig:spectra} (b). In \cref{fig:spectra} (c), $U$ affects the binding energies significantly. This means the energies involved in HHG from impurities are lowered, compared to \cref{fig:spectra} (b), while the higher $U$ means more energy is required to produce a doublon. This explains the trend towards higher gain for lower $\Delta$ values. The overall lower gain in the high $\Delta$ cases, compared to the $\Delta=0.01t_0$ result is likely a result of the much fewer sites which facilitates the impurity-based HHG compared to the doublon-holon-based HHG. HHG from impurities can only originate from sites with binding impurities whereas doublon-holon pairs can be created at any point in the lattice. Fewer generating sites tend to mean less HHG.

	\subsection{Symmetry analysis and populations in states with  different $K$ values}
	In this section, we discuss the shift symmetry discussed in \cref{sec:symmetries}. In \cref{fig:rhos} the population in states with each value of $K_i$, $i\in\left\{0,1,2,\cdots,11\right\}$ is plotted. For $\Delta=0.01t_0$, shown in Fig.~\ref{fig:rhos} (a), (b), and (c), the large majority of the electron population are in states with either $K=0$ or $K=6\times2\pi/La=\pi/a$, which is consistent with the values mentioned for the ground state in \cref{sec:symmetries}. For $\Delta>0$ the ground state cannot be expected to be dominated by a particular $K$ value, since $K$ is no longer a good quantum number. However, as $\Delta\ll t_0$ in Figs.~\ref{fig:rhos} (a), (b), and (c) the Hubbard Hamiltonian, $\hat{H}_\text{Hub}(t)$, dominates the dynamics, and causes the ground state to be predominantly in the $K=0$ or $K=\pi/a$ states. As $\Delta$ increases to $t_0$ in Figs.~\ref{fig:rhos} (d), (e), and (f), and further to $5t_0$ in Figs.~\ref{fig:rhos} (g), (h), and (i) the $K=0$ and $K=\pi/a$ configurations become decreasingly populated in the ground state. As $\hat{H}_\Delta$ couples states with different $K$ values, the observed more evenly distributed population is to be expected. As the $U$ value increases across Figs.~\ref{fig:rhos} (g), $U=0$, (h), $U=5t_0$, and (i), $U=10t_0$, the states become increasingly dominated by the $K=0$ and $K=\pi/a$ states, as is to be expected since increasing $U$ corresponds to increasing the importance of $\hat{H}_U$ relative to $\hat{H}_\Delta$. In \cref{fig:rhos}, a black patch has been added to the colorbar, centered around a population of $1/L$, which corresponds to the electron population being evenly distributed across all $K$ values, and a complete delocalization in $K$-space. One can see from \cref{fig:rhos} that when $\Delta$ dominates but $U$ is close enough to $\Delta$ so that the external laser pulse can bridge the $\Delta$-induced energy gaps, then the pulse drives the system into a state with, at least approximately, evenly distributed population across the different $K$ values, see Figs.~\ref{fig:rhos} (d), (e), and (i). This is in line with the qualitative picture presented in \cref{sec:Qualitative physical picture}. For low $\Delta$, we observe conservation of $K$ regardless of the $U$ value. Conservation of $K$ for $\Delta=0$ is analytically derived and this is, therefore, no surprise. As $\Delta$ increases we observe even population distribution across the possible $K$ values, which is only possible in a system with significant $\Delta$, and the larger $U$ is, the larger $\Delta$ has to be in order to cause even distribution of population across the $K$ values.
	
	Note also from Figs.~\ref{fig:rhos} (d), (e), (h), and (i) how the pulse can drive the population towards delocalization in $K$ space. Finally, we note that this delocalization in $K$ space is quite reminiscent of how real-space localization looks in Bloch space, and might therefore indicate that at least parts of the energy spectra are many-body localized. It seems therefore that the pulse may drive systems that are on the cusp of being many-body localized into a many-body localized subspace of the system.
	
	\subsection{Populations at individual sites}
	\begin{figure*}
		\includegraphics[width=.95\linewidth]{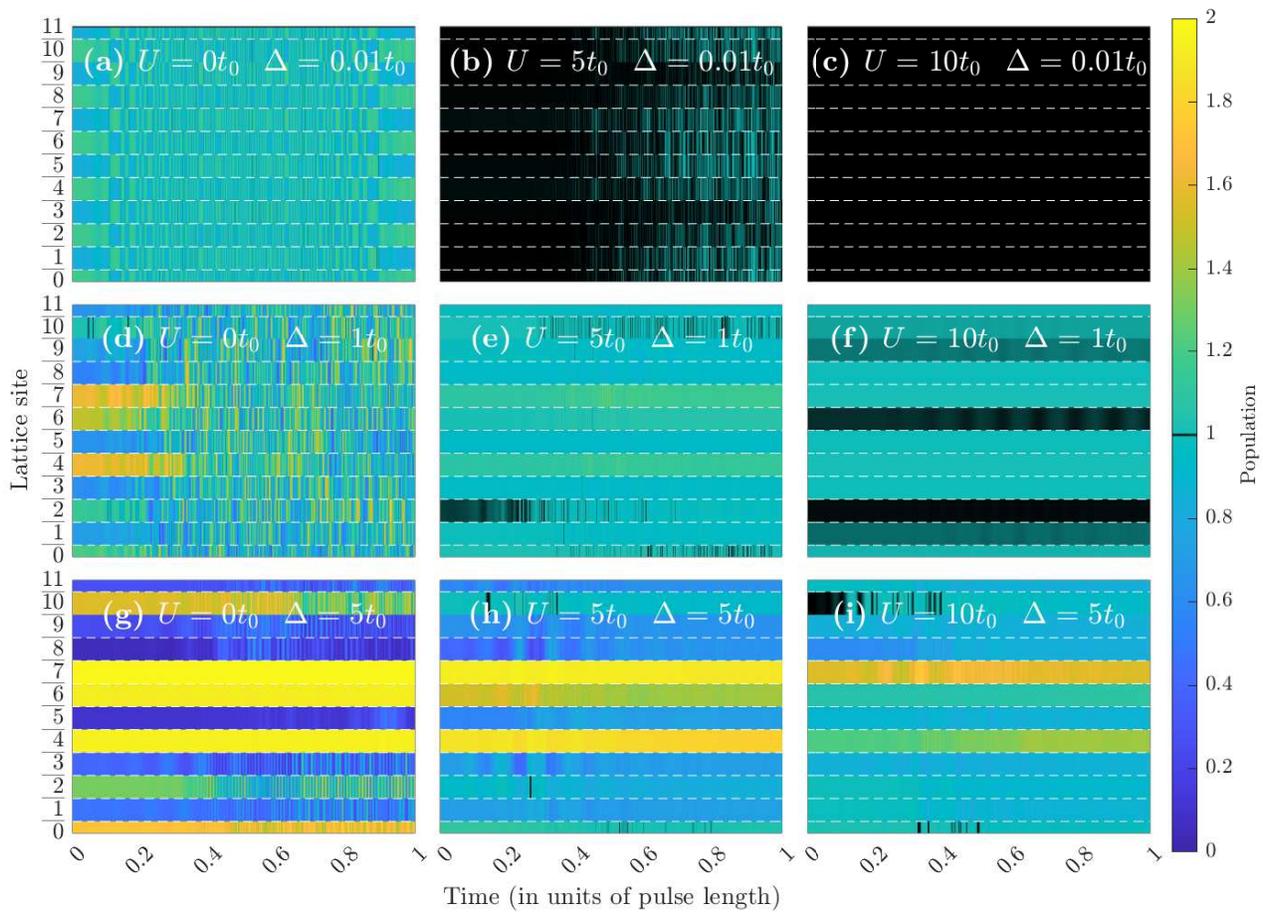}
		\caption{Populations on the different lattice sites. The panels show results for three $\Delta$ values, $\Delta=0.01t_0$ in (a), (b), and (c), $\Delta=t_0$ in (d), (e), and (f), and $\Delta=5t_0$ in (g), (h), and (i) and results for three different $U$ values, $U=0$ in (a), (d), and (g), $U=5t_0$ in (b), (e), and (h), and $U=10t_0$ in (c), (f), and (i). The colorbar has a black patch centered at 1 which corresponds to equal distribution of the population across lattice sites. This black color eases the identification of situations with equal populations on all lattice sites.}
		\label{fig:pops}
	\end{figure*}
	Now, having discussed the populations in each of the sub-spaces with different total crystal momentum we proceed by discussing the populations at the different lattice sites. Those populations are plotted in \cref{fig:pops}. For $\Delta=0.01t_0$ plotted in Figs.~\ref{fig:pops} (a), (b), and (c) where $U=0$, $U=5t_0$, and $U=10t_0$, respectively, increasing $U$ results in an increasingly equal electron distribution across the lattice sites. This is as expected, the high $U$ causing the electrons to repel one another resulting in one electron per lattice site, i.e., the Mott-insulator limit. In \cref{fig:pops} (a) where $U=0$ and $\Delta=0.01t_0$, the population oscillates across the lattice with a periodicity of 2 sites, this is inline with about half the population being in states with $K=0$ and the other being in states with $K=\pi/a$. Summing configurations with those $K$-values will result in a population curve that oscillates with a period of 2 sites. As $\Delta$ increases to $t_0$ in Figs.~\ref{fig:pops} (d), (e), and (f), where $U=0$, $U=5t_0$, and $U=10t_0$, respectively, the populations begin to differ noticeably across the different lattice sites. The potential used for these simulations is the same as is plotted in \cref{fig:illustration}, though scaled with different $\Delta$ values. One can identify the impact of this potential in the initial population of \cref{fig:pops} (d). The lowest points in the potential are, in order, sites 7, 4, and 6, see \cref{fig:illustration}. The populations in the initial ground state, shown in \cref{fig:pops} (d), have high population on those sites, and relatively low population on the others. As $U$ increases so that $U=1t_0$ in \cref{fig:pops} (e) and $U=10t_0$ in \cref{fig:pops} (f) the populations across the different lattice sites become increasingly equal but still with hints of the underlying potential influencing the electron populations at the lattice sites. In Figs.~\ref{fig:pops} (g), (h), and (i) $\Delta=5t_0$ and the potential has significant influence on the populations throughout the simulation. In \cref{fig:pops} (g), $U=0$, the populations are dominated by $\hat{H}_\Delta$	as seen by the Pauli limited populations on sites 4, 6, and 7 and low populations on sites 8, 9, and 11, where the $\Delta$-potential is highest. 
	
	\subsection{Fidelity measure}
\begin{figure}
		\includegraphics[width=\linewidth]{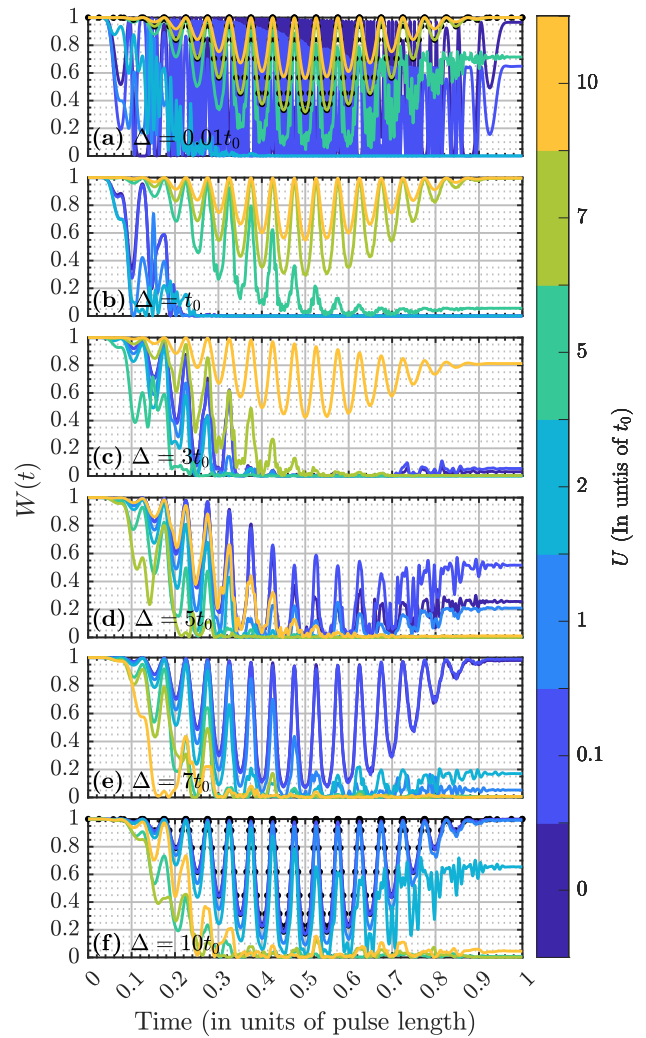}
		\caption{$W(t)$ [\cref{eq:W}] for various values of $U$ and $\Delta$. The $\Delta$ values used are (a)  $\Delta=0.01t_0$, (b) $\Delta=t_0$, (c) $\Delta=3t_0$, (d) $\Delta=5t_0$, (e) $\Delta=7t_0$, and (f) $\Delta=10t_0$. The black dots behind the $U=7t_0$ result in panel (a) and the $U=0.1t_0$ result in panel (f) represent the results obtained from the adiabatic approximation, see \cref{sec: Adiabatic dynamics}.}
		\label{fig:W}
	\end{figure}

	The fidelity measure, $W(t)$ in \cref{eq:W}, is shown in \cref{fig:W} for a variety of $U$ and $\Delta$ values. Starting from \cref{fig:W} (a) $\Delta=0.01t_0$, we observe that for $U=0$, the simulations end in the ground state, which is to be expected as the dynamics are largely given via the following modification to the crystal momenta, $k\rightarrow k+A(t)$. Since the laser pulse starts and ends with zero vector potential, the crystal momentum of each particle ends back in its initial value. In Figs.~\ref{fig:W} (a), $\Delta=0.01t_0$, (b), $\Delta=1t_0$, and (c), $\Delta=3t_0$ the $U=10t_0$ result is closest to unity at the end of the pulse.	We link this behavior to adiabatic dynamics, as described in \cref{sec: Adiabatic dynamics}. The point is that $U=10 t_0$ is much larger than $\Delta$ in these panels, and therefore $U=10 t_0$ sets the energy and time scale. In Fig.~\ref{fig:W} (d), $\Delta =5 t_0$, the $\Delta$ value is such that the considered $U$ values are not sufficiently large to enter the regime of adiabatic dynamics, and $\Delta$ is not sufficiently large to set the energy and time scale, so we observe a large drop in $W(t)$. Finally, in Figs.~\ref{fig:W} (e) and (f), $\Delta$ is much larger and sets the energy and time scale for $U$ up to around $2 t_0$. This means that we again observe adiabatic dynamics , but now for the smallest $U$. We verify that adiabatic behavior is observed by plotting the $W(t)$ obtained in the adiabatic approximation as black dots under the $U=7t_0$, $\Delta=0.01t_0$ case of \cref{fig:W} (a) and the $U=0.1$, $\Delta=10t_0$ case of \cref{fig:W} (f). See also \cref{fig:WU01D10,fig:WU7D001} in \cref{sec: Adiabatic dynamics}. The adiabatic approximation assumes that the rate of change of the Hamiltonian, in this case, due to the presence of the external laser field, is much smaller than the natural time scale of the system. Intuitively adiabatic dynamics make sense in the $\left|U-\Delta\right|\gg 0$ limit as the ground state is dominated either by antiferromagnetically ordered electron configurations or by configurations that place the electron on the sites with lowest $A_i$ values (see Fig.~\ref{fig:illustration}). Such configurations only allow transitions that require a substantial amount of energy. In the $U\gg\Delta$ case, there are transitions that create a doublon-holon pair, which requires an energy of about $U$. In the $\Delta\gg U$ case, the electrons need to leave the $\hat{H}_\Delta$-induced energy well, which requires energy on the order of  $\Delta$. As $\Delta$ increases across Figs.~\ref{fig:W} (b) and (c), the low $U$ results decrease to approximately zero quite quickly and remain close to zero throughout the remainder of the simulation. This is likely due to the low $\Delta$ and, or $U$ breaking the degeneracy of the eigenstates in the Bloch states which makes it easier for the laser pulse to drive the electrons out of the ground state and thereby reduce $W(t)$. As $\Delta$ begins to approach the largest simulated $U$ values in Fig.~\ref{fig:W} (e) $\Delta=7t_0$ and (f) $\Delta=10t_0$, we observe that the results for the highest $U$ values start behaving in the same non-adiabatic form as described earlier. This is in line with the intuitive picture described in \cref{sec:Qualitative physical picture}. When either $\Delta\gg U$ or $U\gg\Delta$, the energy gaps between different electron configurations can be massive, but when $U\simeq \Delta$ they can balance each other out resulting in small, and therefore easily driven, energy gaps between different configurations, see \cref{fig:illustration}. This picture is further validated by the adiabatic behavior of the low $U$ cases in the highest $\Delta$ cases. Finally, the results for  the largest $\Delta$ values in Figs.~\ref{fig:W} (e) and (f), $U=7t_0$ and $U=10t_0$, respectively, qualitatively behave like the low $\Delta$ results of Figs.~\ref{fig:W} (b), $U=t_0$ and (c) $\Delta=3t_0$, likely because of the comparable $U$ and $\Delta$ values.
	 
	\subsection{Averages over spectra}
	\begin{figure}
		\includegraphics[width=\linewidth]{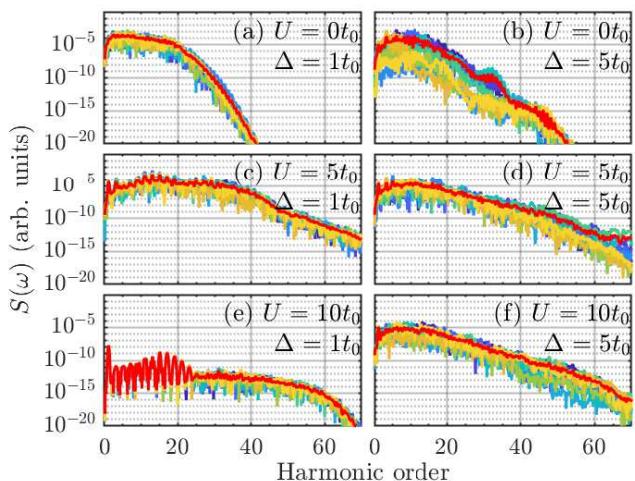}
		\caption{Spectra generated from ten different random realizations of $A_i$-values, in accordance with the probability distribution described in connection with \cref{eq:Ai}. Such a set of spectra is generated for a variety of $U$ and $\Delta$ values, (a) and (b), $U=0$, (c) and (d), $U=5t_0$, and (e) and (f) $U=10t_0$. The $\Delta$ values in (a), (c), and (e) are $\Delta=1t_0$, and in (b), (d), and (f) $\Delta=5t_0$. The red spectra are the averages of the other spectra. Each set of $A_i$ values corresponds to a given color.}
		\label{fig:mean spec}
	\end{figure} 
	
	In order to discuss and describe the effects of $\hat{H}_\Delta$ on the spectra in a more general sense, we have calculated the spectra for ten different random sets of $A_i$-values, $\left\{A_i\right\}$, each generated according to \cref{eq:Ai}. Those spectra are plotted in \cref{fig:mean spec}. In \cref{fig:mean spec} (a), $U=0$ and $\Delta=t_0$, we observe two noteworthy things. Firstly, there are only relatively small differences between the spectra for different $\left\{A_i\right\}$ sets, which makes sense as $\Delta$ remains relatively small. Secondly, the general structure of the spectra, particularly the averaged spectra, consists of a plateau that ends at around the 20th harmonic order followed by a dropoff. The bandwidth of the Bloch band plus 2 times the $\Delta$ value used here corresponds to about 23 harmonic orders. The 20 harmonic orders are therefore close to the maximal possible energy transition. As the probability of having picked $A_i$ values which facilitate the maximal possible energy transition is rather low it seems that the energy around 20 harmonic orders is the maximal energy associated with a transition in the lattice for a given random set of $\left\{A_i\right\}$. This explains the overall spectral structure. As $\Delta$ increases, see \cref{fig:mean spec} (b) where $U=0$ and $\Delta=5t_0$, the spectral dependence on the $A_i$ values increases. This is seen from the substantially larger difference between the spectra from different sets of $A_i$-values. This dependence indicates that each spectrum depends significantly on the specific energy gaps for the individual realizations of the $A_i$ values and the ability of the pulse to drive electrons past those energy gaps, or lack of ability to do so.  For larger $U$ values, see \cref{fig:mean spec} (c) and (d) $U=5t_0$ with $\Delta=t_0$ and $\Delta=5t_0$, respectively, we observe less of a dependence on the specific $A_i$ values. This is in line with the earlier presented qualitative picture, in the sense that the effect of the $\Delta$ and $U$ terms tend to balance each other out. 
	
	The discussion in this section does not alter the conclusions drawn earlier regarding the behavior of the spectra as $U$ and $\Delta$ change, the validity of the qualitative picture described in \cref{sec:Qualitative physical picture}, the symmetry breaking around \cref{fig:rhos}, nor the dynamics described around \cref{fig:W}.
	
 	\section{Summary and conclusion}\label{sec:Conclusion}
 	In this paper, we used the Hubbard model to study the effects of lattice imperfections on HHG from correlated systems. We simulated lattice imperfections by adding a random modification to the chemical potential at each lattice site which is in line with the procedure taken in studies of Anderson localization in the uncorrelated limit. Previous studies of HHG from highly correlated systems have, as far as we are aware, all operated under the assumption of a perfect lattice. Such studies have led to the doublon-holon-based three-step model \cite{Murakami2021} and similar arguments largely based around correlation-induced energy transitions \cite{Murakami2021, Hansen22_2}. As discussed here, lattice imperfections can modify the dynamics of such a system significantly and are therefore interesting to include in the simulation. Beyond that, the perfect lattice assumption leads to a shift symmetry in the lattice, which is unlikely to hold in real solids and which is broken by the random lattice noise term. We found it interesting to study the effects of breaking that system symmetry. 
 	
 	In this work, we set out to answer the following questions.  (i) How are the HHG spectra affected by the addition of lattice imperfections for non-vanishing $U$? (ii) How do the effects of lattice imperfections and electron-electron correlation interact? and (iii) How do the dynamics in the lattice relate to the spectra? We answered these questions by first introducing a qualitative picture of the system under the effects of both correlation and imperfections. This picture bases itself on the idea that the lattice imperfections lead to ground states with high electron populations on a few sites, those with the lowest chemical potential, while correlation results in an even population distribution throughout the lattice. This fundamental difference between the effects induced by imperfections and correlations results in an internal struggle between the two effects, resulting in more dynamics when $\Delta\approx U$. This picture was verified using, among other things, the harmonic spectra. The spectra showed that when there was little to no correlation, the effects of imperfections are largely the same as those of correlation. As soon as imperfections are added the spectra show significant gain in the medium to high harmonic orders, but less gain for the lowest orders. This continues until the $\Delta$ value becomes very large, at which point the overall spectra begin to lose gain as the electrons become locked at the sites with the lowest chemical potential. As correlation was added on top of the imperfections the effects of imperfections largely disappeared, so long as the correlation remained small. This is likely because the effects on the spectra of imperfections and correlations are largely the same so long as they both remain relatively small, so the effects of adding a relatively small correlation to a small $\Delta$ are not particularly significant. As $U$ approaches $U=10t_0$, the effects of the imperfections become increasingly noticeable. Severe imperfections result in increased harmonic gain among the lower harmonic, opposite of the effect when correlation was low. We link this to the balancing act described earlier. When correlation and the imperfections balance each other that balance facilitates low-energy transitions, which are easy for the pulse to drive and result in low-order harmonics. 
 	
	We then proceeded to discuss the shift symmetry, which the lattice imperfection breaks. Without imperfection the system is invariant under shift or translations of the whole system by integer site index. We found that, as imperfections were added, the electrons end up distributed evenly across all possible values of the total crystal momentum. This happened in a manner similar to localization in correlation-free Bloch space. 
 	
 	We then proceeded to discuss the fidelity measure, i.e., the population in the initial state as a function of time. We found that in the high-correlation, low-imperfection limit, or the opposite limit of severe imperfections but little to no correlation, the dynamics are largely adiabatic in nature. This follows as the energy splittings between the states are quite large in those limits. However, in every case where the balancing act was in effect, we found that the fidelity measure decayed much more severely, likely a result of the low-energy transitions which the qualitative picture predicts in the cases of balanced imperfections and correlation. Finally, we discussed the effects of picking different values for the random modifications to the chemical potential. We found that as long as the imperfections are relatively small, or merely small compared to the correlation level, the precise choice of modifications has a relatively small impact and that the impact increases with the severity of the imperfections. We do also conclude that the trends described above should hold for all reasonable choices of modifications. 
 	
 	In short, the effects of lattice imperfections on the spectra seem largely similar to those of correlation, with the interesting add-on of correlation and the imperfections balancing each other resulting in Bloch-like transitions in lattices with very large correlation and imperfections. So if the correlation is low then imperfections result in increased high harmonic gain, whereas if the correlation is large then imperfections result in gain in the lower order harmonic orders.

 	 \acknowledgements
 	This work was supported by the Independent Research Fund Denmark (Grant No. 1026-00040B).
 	\appendix
 	\section{Adiabatic dynamics}\label{sec: Adiabatic dynamics}
 	In this appendix, we give some details of the adiabatic results shown in Fig.~\ref{fig:W} of the main text.
 	If the rate at which the external laser field changes is much slower than the natural timescale of the dynamics in said system, then it seems intuitively reasonable to assume that the quantum state of the system merely evolves with the system, without transitioning from one eigenstate to another. In Ref.~\cite{GriffithsQuant} it is derived that assuming no degeneracy
 	\begin{widetext}
 	\begin{align}
 		\dot{c}_m(t)&=-c_m\bra{\psi_m}\ket{ \dot{\psi}_m}-\sum_{n \neq m} c_n \frac{\bra{\psi_m}\dot{\hat{H}}(t) \ket{\psi_n}}{E_n(t)-E_m(t)} e^{-i  \int_0^t\left[E_n\left(t^{\prime}\right)-E_m\left(t^{\prime}\right)\right] d t^{\prime}}\label{eq:Adia1}
 	\end{align}
	\end{widetext}
	where $c_m(t)=\bra{\psi_m(t)}\ket{\Psi(t)}$, with $\ket{\psi_m(t)}$ being the $m$'th adiabatic eigenstate of the Hamiltonian at the instant $t$, and $\ket{\Psi(t)}$ being the state of the system at time $t$, $\hat{H}(t)$ being the Hamiltonian of the system and $E_m(t)$ being the eigenvalue of $\ket{\psi_m(t)}$, finally time derivatives are denoted by a dot. The adiabatic approximation then consists of assuming that $\forall n,m: \left|\frac{\bra{\psi_m}\dot{\hat{H}}(t) \ket{\psi_n}}{E_n(t)-E_m(t)}\right|\ll 1$, resulting in the neglect of the second term in \cref{eq:Adia1} and the following solution
	\begin{align}
			c_m(t)&=c_m(0)e^{-\int_{0}^{t}\bra{\psi_m(t')}\ket{ \dot{\psi}_m(t')}dt'},
	\end{align}
	Thus when the system changes adiabatically the state of the system follows the continuous change of the eigenstate. If the system starts in the ground state then it will remain in the time-dependent instantaneous ground state of the system at all times. This makes calculating the $W$-measure under the adiabatic approximation relatively straightforward, simply find the ground state of the Hamiltonian of Eq.~(\ref{eq: Hamiltonian}) at various times, and project it onto the field-free ground state. 
	
\begin{figure}[h]
		\includegraphics[width=\linewidth]{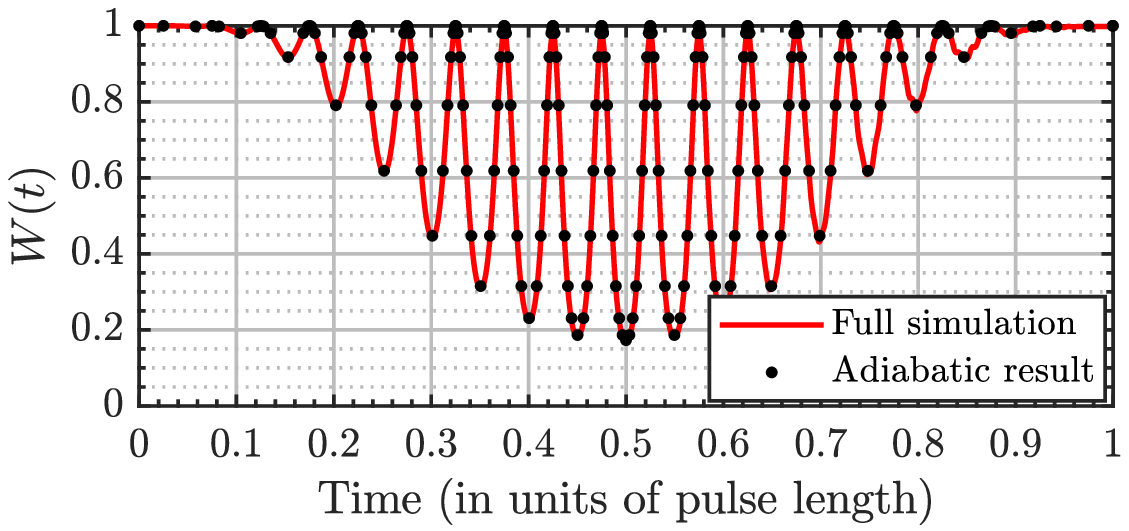}
		\caption{$W(t)$ for the $U=0.1t_0$ and $\Delta=10t_0$ results. The red line was obtained using the full simulation, while the black dots are from the adiabatic approximation. These results are also presented in \cref{fig:W} (f).} 
		\label{fig:WU01D10}
	\end{figure}

\begin{figure}[H]
	\includegraphics[width=\linewidth]{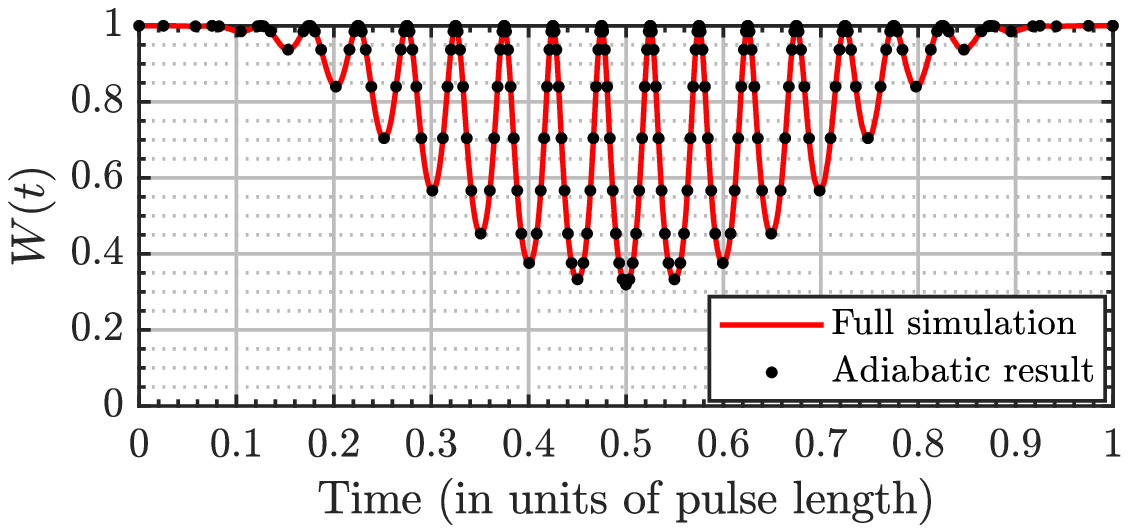}
	\caption{$W(t)$ for the $U=7t_0$ and $\Delta=0.01t_0$ results. The red line was obtained using the full simulation, while the black dots are from the adiabatic approximation. These results are also presented in \cref{fig:W} (a).}
	\label{fig:WU7D001}
\end{figure}

	To test how well this approximation was in the $\Delta\gg \left(U,\omega_L\right)$ or $U\gg\left(\Delta,\omega_L\right)$ cases we calculated the adiabatic approximation for the fidelity measure in the cases of $U=0.1t_0$, $\Delta=10t_0$, see \cref{fig:WU01D10} and $U=7t_0$, $\Delta=0.01t_0$ \cref{fig:WU7D001}. In both cases, there is close to perfect agreement between the exact result and the adiabatic approximation.

\bibliography{sources_self3.bib}

\begin{thebibliography}{62}%
\makeatletter
\providecommand \@ifxundefined [1]{%
 \@ifx{#1\undefined}
}%
\providecommand \@ifnum [1]{%
 \ifnum #1\expandafter \@firstoftwo
 \else \expandafter \@secondoftwo
 \fi
}%
\providecommand \@ifx [1]{%
 \ifx #1\expandafter \@firstoftwo
 \else \expandafter \@secondoftwo
 \fi
}%
\providecommand \natexlab [1]{#1}%
\providecommand \enquote  [1]{``#1''}%
\providecommand \bibnamefont  [1]{#1}%
\providecommand \bibfnamefont [1]{#1}%
\providecommand \citenamefont [1]{#1}%
\providecommand \href@noop [0]{\@secondoftwo}%
\providecommand \href [0]{\begingroup \@sanitize@url \@href}%
\providecommand \@href[1]{\@@startlink{#1}\@@href}%
\providecommand \@@href[1]{\endgroup#1\@@endlink}%
\providecommand \@sanitize@url [0]{\catcode `\\12\catcode `\$12\catcode
  `\&12\catcode `\#12\catcode `\^12\catcode `\_12\catcode `\%12\relax}%
\providecommand \@@startlink[1]{}%
\providecommand \@@endlink[0]{}%
\providecommand \url  [0]{\begingroup\@sanitize@url \@url }%
\providecommand \@url [1]{\endgroup\@href {#1}{\urlprefix }}%
\providecommand \urlprefix  [0]{URL }%
\providecommand \Eprint [0]{\href }%
\providecommand \doibase [0]{https://doi.org/}%
\providecommand \selectlanguage [0]{\@gobble}%
\providecommand \bibinfo  [0]{\@secondoftwo}%
\providecommand \bibfield  [0]{\@secondoftwo}%
\providecommand \translation [1]{[#1]}%
\providecommand \BibitemOpen [0]{}%
\providecommand \bibitemStop [0]{}%
\providecommand \bibitemNoStop [0]{.\EOS\space}%
\providecommand \EOS [0]{\spacefactor3000\relax}%
\providecommand \BibitemShut  [1]{\csname bibitem#1\endcsname}%
\let\auto@bib@innerbib\@empty
\bibitem [{\citenamefont {Li}\ \emph {et~al.}(2008)\citenamefont {Li},
  \citenamefont {Zhou}, \citenamefont {Lock}, \citenamefont {Patchkovskii},
  \citenamefont {Stolow}, \citenamefont {Kapteyn},\ and\ \citenamefont
  {Murnane}}]{Li2008}%
  \BibitemOpen
  \bibfield  {author} {\bibinfo {author} {\bibfnamefont {W.}~\bibnamefont
  {Li}}, \bibinfo {author} {\bibfnamefont {X.}~\bibnamefont {Zhou}}, \bibinfo
  {author} {\bibfnamefont {R.}~\bibnamefont {Lock}}, \bibinfo {author}
  {\bibfnamefont {S.}~\bibnamefont {Patchkovskii}}, \bibinfo {author}
  {\bibfnamefont {A.}~\bibnamefont {Stolow}}, \bibinfo {author} {\bibfnamefont
  {H.~C.}\ \bibnamefont {Kapteyn}},\ and\ \bibinfo {author} {\bibfnamefont
  {M.~M.}\ \bibnamefont {Murnane}},\ }\bibfield  {title} {\bibinfo {title}
  {Time-resolved dynamics in probed using high harmonic generation},\ }\href
  {https://doi.org/10.1126/science.1163077} {\bibfield  {journal} {\bibinfo
  {journal} {Science}\ }\textbf {\bibinfo {volume} {322}},\ \bibinfo {pages}
  {1207} (\bibinfo {year} {2008})}\BibitemShut {NoStop}%
\bibitem [{\citenamefont {Lein}\ \emph {et~al.}(2002)\citenamefont {Lein},
  \citenamefont {Hay}, \citenamefont {Velotta}, \citenamefont {Marangos},\ and\
  \citenamefont {Knight}}]{Lein2002}%
  \BibitemOpen
  \bibfield  {author} {\bibinfo {author} {\bibfnamefont {M.}~\bibnamefont
  {Lein}}, \bibinfo {author} {\bibfnamefont {N.}~\bibnamefont {Hay}}, \bibinfo
  {author} {\bibfnamefont {R.}~\bibnamefont {Velotta}}, \bibinfo {author}
  {\bibfnamefont {J.~P.}\ \bibnamefont {Marangos}},\ and\ \bibinfo {author}
  {\bibfnamefont {P.~L.}\ \bibnamefont {Knight}},\ }\bibfield  {title}
  {\bibinfo {title} {Interference effects in high-order harmonic generation
  with molecules},\ }\href {https://doi.org/10.1103/PhysRevA.66.023805}
  {\bibfield  {journal} {\bibinfo  {journal} {Phys. Rev. A}\ }\textbf {\bibinfo
  {volume} {66}},\ \bibinfo {pages} {023805} (\bibinfo {year}
  {2002})}\BibitemShut {NoStop}%
\bibitem [{\citenamefont {Torres}\ \emph {et~al.}(2007)\citenamefont {Torres},
  \citenamefont {Kajumba}, \citenamefont {Underwood}, \citenamefont {Robinson},
  \citenamefont {Baker}, \citenamefont {Tisch}, \citenamefont {de~Nalda},
  \citenamefont {Bryan}, \citenamefont {Velotta}, \citenamefont {Altucci},
  \citenamefont {Turcu},\ and\ \citenamefont {Marangos}}]{Torres2007}%
  \BibitemOpen
  \bibfield  {author} {\bibinfo {author} {\bibfnamefont {R.}~\bibnamefont
  {Torres}}, \bibinfo {author} {\bibfnamefont {N.}~\bibnamefont {Kajumba}},
  \bibinfo {author} {\bibfnamefont {J.~G.}\ \bibnamefont {Underwood}}, \bibinfo
  {author} {\bibfnamefont {J.~S.}\ \bibnamefont {Robinson}}, \bibinfo {author}
  {\bibfnamefont {S.}~\bibnamefont {Baker}}, \bibinfo {author} {\bibfnamefont
  {J.~W.~G.}\ \bibnamefont {Tisch}}, \bibinfo {author} {\bibfnamefont
  {R.}~\bibnamefont {de~Nalda}}, \bibinfo {author} {\bibfnamefont {W.~A.}\
  \bibnamefont {Bryan}}, \bibinfo {author} {\bibfnamefont {R.}~\bibnamefont
  {Velotta}}, \bibinfo {author} {\bibfnamefont {C.}~\bibnamefont {Altucci}},
  \bibinfo {author} {\bibfnamefont {I.~C.~E.}\ \bibnamefont {Turcu}},\ and\
  \bibinfo {author} {\bibfnamefont {J.~P.}\ \bibnamefont {Marangos}},\
  }\bibfield  {title} {\bibinfo {title} {Probing orbital structure of
  polyatomic molecules by high-order harmonic generation},\ }\href
  {https://doi.org/10.1103/PhysRevLett.98.203007} {\bibfield  {journal}
  {\bibinfo  {journal} {Phys. Rev. Lett.}\ }\textbf {\bibinfo {volume} {98}},\
  \bibinfo {pages} {203007} (\bibinfo {year} {2007})}\BibitemShut {NoStop}%
\bibitem [{\citenamefont {Kraus}\ \emph {et~al.}(2015)\citenamefont {Kraus},
  \citenamefont {Mignolet}, \citenamefont {Baykusheva}, \citenamefont
  {Rupenyan}, \citenamefont {Horný}, \citenamefont {Penka}, \citenamefont
  {Grassi}, \citenamefont {Tolstikhin}, \citenamefont {Schneider},
  \citenamefont {Jensen}, \citenamefont {Madsen}, \citenamefont {Bandrauk},
  \citenamefont {Remacle},\ and\ \citenamefont {Wörner}}]{Kraus2015}%
  \BibitemOpen
  \bibfield  {author} {\bibinfo {author} {\bibfnamefont {P.~M.}\ \bibnamefont
  {Kraus}}, \bibinfo {author} {\bibfnamefont {B.}~\bibnamefont {Mignolet}},
  \bibinfo {author} {\bibfnamefont {D.}~\bibnamefont {Baykusheva}}, \bibinfo
  {author} {\bibfnamefont {A.}~\bibnamefont {Rupenyan}}, \bibinfo {author}
  {\bibfnamefont {L.}~\bibnamefont {Horný}}, \bibinfo {author} {\bibfnamefont
  {E.~F.}\ \bibnamefont {Penka}}, \bibinfo {author} {\bibfnamefont
  {G.}~\bibnamefont {Grassi}}, \bibinfo {author} {\bibfnamefont {O.~I.}\
  \bibnamefont {Tolstikhin}}, \bibinfo {author} {\bibfnamefont
  {J.}~\bibnamefont {Schneider}}, \bibinfo {author} {\bibfnamefont
  {F.}~\bibnamefont {Jensen}}, \bibinfo {author} {\bibfnamefont {L.~B.}\
  \bibnamefont {Madsen}}, \bibinfo {author} {\bibfnamefont {A.~D.}\
  \bibnamefont {Bandrauk}}, \bibinfo {author} {\bibfnamefont {F.}~\bibnamefont
  {Remacle}},\ and\ \bibinfo {author} {\bibfnamefont {H.~J.}\ \bibnamefont
  {Wörner}},\ }\bibfield  {title} {\bibinfo {title} {Measurement and laser
  control of attosecond charge migration in ionized iodoacetylene},\ }\href
  {https://doi.org/10.1126/science.aab2160} {\bibfield  {journal} {\bibinfo
  {journal} {Science}\ }\textbf {\bibinfo {volume} {350}},\ \bibinfo {pages}
  {790} (\bibinfo {year} {2015})}\BibitemShut {NoStop}%
\bibitem [{\citenamefont {Luu}\ and\ \citenamefont
  {W{\"o}rner}(2018)}]{Luu2018}%
  \BibitemOpen
  \bibfield  {author} {\bibinfo {author} {\bibfnamefont {T.~T.}\ \bibnamefont
  {Luu}}\ and\ \bibinfo {author} {\bibfnamefont {H.~J.}\ \bibnamefont
  {W{\"o}rner}},\ }\bibfield  {title} {\bibinfo {title} {Measurement of the
  {{B}}erry curvature of solids using high-harmonic spectroscopy},\ }\href
  {https://doi.org/10.1038/s41467-018-03397-4} {\bibfield  {journal} {\bibinfo
  {journal} {Nature Communications}\ }\textbf {\bibinfo {volume} {9}},\
  \bibinfo {pages} {916} (\bibinfo {year} {2018})}\BibitemShut {NoStop}%
\bibitem [{\citenamefont {Silva}\ \emph {et~al.}(2018)\citenamefont {Silva},
  \citenamefont {Blinov}, \citenamefont {Rubtsov}, \citenamefont {Smirnova},\
  and\ \citenamefont {Ivanov}}]{Silva2018}%
  \BibitemOpen
  \bibfield  {author} {\bibinfo {author} {\bibfnamefont {R.~E.~F.}\
  \bibnamefont {Silva}}, \bibinfo {author} {\bibfnamefont {I.~V.}\ \bibnamefont
  {Blinov}}, \bibinfo {author} {\bibfnamefont {A.~N.}\ \bibnamefont {Rubtsov}},
  \bibinfo {author} {\bibfnamefont {O.}~\bibnamefont {Smirnova}},\ and\
  \bibinfo {author} {\bibfnamefont {M.}~\bibnamefont {Ivanov}},\ }\bibfield
  {title} {\bibinfo {title} {High-harmonic spectroscopy of ultrafast many-body
  dynamics in strongly correlated systems},\ }\href
  {https://doi.org/10.1038/s41566-018-0129-0} {\bibfield  {journal} {\bibinfo
  {journal} {Nature Photonics}\ }\textbf {\bibinfo {volume} {12}},\ \bibinfo
  {pages} {266} (\bibinfo {year} {2018})}\BibitemShut {NoStop}%
\bibitem [{\citenamefont {Schubert}\ \emph {et~al.}(2014)\citenamefont
  {Schubert}, \citenamefont {Hohenleutner}, \citenamefont {Langer},
  \citenamefont {Urbanek}, \citenamefont {Lange}, \citenamefont {Huttner},
  \citenamefont {Golde}, \citenamefont {Meier}, \citenamefont {Kira},
  \citenamefont {Koch},\ and\ \citenamefont {Huber}}]{Schubert2014}%
  \BibitemOpen
  \bibfield  {author} {\bibinfo {author} {\bibfnamefont {O.}~\bibnamefont
  {Schubert}}, \bibinfo {author} {\bibfnamefont {M.}~\bibnamefont
  {Hohenleutner}}, \bibinfo {author} {\bibfnamefont {F.}~\bibnamefont
  {Langer}}, \bibinfo {author} {\bibfnamefont {B.}~\bibnamefont {Urbanek}},
  \bibinfo {author} {\bibfnamefont {C.}~\bibnamefont {Lange}}, \bibinfo
  {author} {\bibfnamefont {U.}~\bibnamefont {Huttner}}, \bibinfo {author}
  {\bibfnamefont {D.}~\bibnamefont {Golde}}, \bibinfo {author} {\bibfnamefont
  {T.}~\bibnamefont {Meier}}, \bibinfo {author} {\bibfnamefont
  {M.}~\bibnamefont {Kira}}, \bibinfo {author} {\bibfnamefont {S.~W.}\
  \bibnamefont {Koch}},\ and\ \bibinfo {author} {\bibfnamefont
  {R.}~\bibnamefont {Huber}},\ }\bibfield  {title} {\bibinfo {title} {Sub-cycle
  control of terahertz high-harmonic generation by dynamical $\text{B}$loch
  oscillations},\ }\href {https://doi.org/10.1038/nphoton.2013.349} {\bibfield
  {journal} {\bibinfo  {journal} {Nature Photonics}\ }\textbf {\bibinfo
  {volume} {8}},\ \bibinfo {pages} {119} (\bibinfo {year} {2014})}\BibitemShut
  {NoStop}%
\bibitem [{\citenamefont {Lewenstein}\ \emph {et~al.}(1994)\citenamefont
  {Lewenstein}, \citenamefont {Balcou}, \citenamefont {Ivanov}, \citenamefont
  {L'Huillier},\ and\ \citenamefont {Corkum}}]{Lewenstein1994}%
  \BibitemOpen
  \bibfield  {author} {\bibinfo {author} {\bibfnamefont {M.}~\bibnamefont
  {Lewenstein}}, \bibinfo {author} {\bibfnamefont {P.}~\bibnamefont {Balcou}},
  \bibinfo {author} {\bibfnamefont {M.~Y.}\ \bibnamefont {Ivanov}}, \bibinfo
  {author} {\bibfnamefont {A.}~\bibnamefont {L'Huillier}},\ and\ \bibinfo
  {author} {\bibfnamefont {P.~B.}\ \bibnamefont {Corkum}},\ }\bibfield  {title}
  {\bibinfo {title} {Theory of high-harmonic generation by low-frequency laser
  fields},\ }\href {https://doi.org/10.1103/PhysRevA.49.2117} {\bibfield
  {journal} {\bibinfo  {journal} {Phys. Rev. A}\ }\textbf {\bibinfo {volume}
  {49}},\ \bibinfo {pages} {2117} (\bibinfo {year} {1994})}\BibitemShut
  {NoStop}%
\bibitem [{\citenamefont {Lein}\ and\ \citenamefont {Rost}(2003)}]{Lein2003}%
  \BibitemOpen
  \bibfield  {author} {\bibinfo {author} {\bibfnamefont {M.}~\bibnamefont
  {Lein}}\ and\ \bibinfo {author} {\bibfnamefont {J.~M.}\ \bibnamefont
  {Rost}},\ }\bibfield  {title} {\bibinfo {title} {Ultrahigh harmonics from
  laser-assisted ion-atom collisions},\ }\href
  {https://doi.org/10.1103/PhysRevLett.91.243901} {\bibfield  {journal}
  {\bibinfo  {journal} {Phys. Rev. Lett.}\ }\textbf {\bibinfo {volume} {91}},\
  \bibinfo {pages} {243901} (\bibinfo {year} {2003})}\BibitemShut {NoStop}%
\bibitem [{\citenamefont {Ghimire}\ \emph {et~al.}(2011)\citenamefont
  {Ghimire}, \citenamefont {DiChiara}, \citenamefont {Sistrunk}, \citenamefont
  {Agostini}, \citenamefont {DiMauro},\ and\ \citenamefont
  {Reis}}]{Ghimire2011}%
  \BibitemOpen
  \bibfield  {author} {\bibinfo {author} {\bibfnamefont {S.}~\bibnamefont
  {Ghimire}}, \bibinfo {author} {\bibfnamefont {A.~D.}\ \bibnamefont
  {DiChiara}}, \bibinfo {author} {\bibfnamefont {E.}~\bibnamefont {Sistrunk}},
  \bibinfo {author} {\bibfnamefont {P.}~\bibnamefont {Agostini}}, \bibinfo
  {author} {\bibfnamefont {L.~F.}\ \bibnamefont {DiMauro}},\ and\ \bibinfo
  {author} {\bibfnamefont {D.~A.}\ \bibnamefont {Reis}},\ }\bibfield  {title}
  {\bibinfo {title} {Observation of high-order harmonic generation in a bulk
  crystal},\ }\href {https://doi.org/10.1038/nphys1847} {\bibfield  {journal}
  {\bibinfo  {journal} {Nature Physics}\ }\textbf {\bibinfo {volume} {7}},\
  \bibinfo {pages} {138} (\bibinfo {year} {2011})}\BibitemShut {NoStop}%
\bibitem [{\citenamefont {Garg}\ \emph {et~al.}(2016)\citenamefont {Garg},
  \citenamefont {Zhan}, \citenamefont {Luu}, \citenamefont {Lakhotia},
  \citenamefont {Klostermann}, \citenamefont {Guggenmos},\ and\ \citenamefont
  {Goulielmakis}}]{Garg2016}%
  \BibitemOpen
  \bibfield  {author} {\bibinfo {author} {\bibfnamefont {M.}~\bibnamefont
  {Garg}}, \bibinfo {author} {\bibfnamefont {M.}~\bibnamefont {Zhan}}, \bibinfo
  {author} {\bibfnamefont {T.~T.}\ \bibnamefont {Luu}}, \bibinfo {author}
  {\bibfnamefont {H.}~\bibnamefont {Lakhotia}}, \bibinfo {author}
  {\bibfnamefont {T.}~\bibnamefont {Klostermann}}, \bibinfo {author}
  {\bibfnamefont {A.}~\bibnamefont {Guggenmos}},\ and\ \bibinfo {author}
  {\bibfnamefont {E.}~\bibnamefont {Goulielmakis}},\ }\bibfield  {title}
  {\bibinfo {title} {Multi-petahertz electronic metrology},\ }\href
  {https://doi.org/10.1038/nature19821} {\bibfield  {journal} {\bibinfo
  {journal} {Nature}\ }\textbf {\bibinfo {volume} {538}},\ \bibinfo {pages}
  {359} (\bibinfo {year} {2016})}\BibitemShut {NoStop}%
\bibitem [{\citenamefont {Luu}\ \emph {et~al.}(2015)\citenamefont {Luu},
  \citenamefont {Garg}, \citenamefont {Kruchinin}, \citenamefont {Moulet},
  \citenamefont {Hassan},\ and\ \citenamefont {Goulielmakis}}]{Luu2015}%
  \BibitemOpen
  \bibfield  {author} {\bibinfo {author} {\bibfnamefont {T.~T.}\ \bibnamefont
  {Luu}}, \bibinfo {author} {\bibfnamefont {M.}~\bibnamefont {Garg}}, \bibinfo
  {author} {\bibfnamefont {S.~Y.}\ \bibnamefont {Kruchinin}}, \bibinfo {author}
  {\bibfnamefont {A.}~\bibnamefont {Moulet}}, \bibinfo {author} {\bibfnamefont
  {M.~T.}\ \bibnamefont {Hassan}},\ and\ \bibinfo {author} {\bibfnamefont
  {E.}~\bibnamefont {Goulielmakis}},\ }\bibfield  {title} {\bibinfo {title}
  {Extreme ultraviolet high-harmonic spectroscopy of solids},\ }\href
  {https://doi.org/10.1038/nature14456} {\bibfield  {journal} {\bibinfo
  {journal} {Nature}\ }\textbf {\bibinfo {volume} {521}},\ \bibinfo {pages}
  {498} (\bibinfo {year} {2015})}\BibitemShut {NoStop}%
\bibitem [{\citenamefont {You}\ \emph {et~al.}(2017)\citenamefont {You},
  \citenamefont {Reis},\ and\ \citenamefont {Ghimire}}]{You2017}%
  \BibitemOpen
  \bibfield  {author} {\bibinfo {author} {\bibfnamefont {Y.~S.}\ \bibnamefont
  {You}}, \bibinfo {author} {\bibfnamefont {D.~A.}\ \bibnamefont {Reis}},\ and\
  \bibinfo {author} {\bibfnamefont {S.}~\bibnamefont {Ghimire}},\ }\bibfield
  {title} {\bibinfo {title} {Anisotropic high-harmonic generation in bulk
  crystals},\ }\href {https://doi.org/10.1038/nphys3955} {\bibfield  {journal}
  {\bibinfo  {journal} {Nature Physics}\ }\textbf {\bibinfo {volume} {13}},\
  \bibinfo {pages} {345} (\bibinfo {year} {2017})}\BibitemShut {NoStop}%
\bibitem [{\citenamefont {Kaneshima}\ \emph {et~al.}(2018)\citenamefont
  {Kaneshima}, \citenamefont {Shinohara}, \citenamefont {Takeuchi},
  \citenamefont {Ishii}, \citenamefont {Imasaka}, \citenamefont {Kaji},
  \citenamefont {Ashihara}, \citenamefont {Ishikawa},\ and\ \citenamefont
  {Itatani}}]{Kaneshima2018}%
  \BibitemOpen
  \bibfield  {author} {\bibinfo {author} {\bibfnamefont {K.}~\bibnamefont
  {Kaneshima}}, \bibinfo {author} {\bibfnamefont {Y.}~\bibnamefont
  {Shinohara}}, \bibinfo {author} {\bibfnamefont {K.}~\bibnamefont {Takeuchi}},
  \bibinfo {author} {\bibfnamefont {N.}~\bibnamefont {Ishii}}, \bibinfo
  {author} {\bibfnamefont {K.}~\bibnamefont {Imasaka}}, \bibinfo {author}
  {\bibfnamefont {T.}~\bibnamefont {Kaji}}, \bibinfo {author} {\bibfnamefont
  {S.}~\bibnamefont {Ashihara}}, \bibinfo {author} {\bibfnamefont {K.~L.}\
  \bibnamefont {Ishikawa}},\ and\ \bibinfo {author} {\bibfnamefont
  {J.}~\bibnamefont {Itatani}},\ }\bibfield  {title} {\bibinfo {title}
  {Polarization-resolved study of high harmonics from bulk semiconductors},\
  }\href {https://doi.org/10.1103/PhysRevLett.120.243903} {\bibfield  {journal}
  {\bibinfo  {journal} {Phys. Rev. Lett.}\ }\textbf {\bibinfo {volume} {120}},\
  \bibinfo {pages} {243903} (\bibinfo {year} {2018})}\BibitemShut {NoStop}%
\bibitem [{\citenamefont {Liu}\ \emph {et~al.}(2017)\citenamefont {Liu},
  \citenamefont {Li}, \citenamefont {You}, \citenamefont {Ghimire},
  \citenamefont {Heinz},\ and\ \citenamefont {Reis}}]{Liu2017}%
  \BibitemOpen
  \bibfield  {author} {\bibinfo {author} {\bibfnamefont {H.}~\bibnamefont
  {Liu}}, \bibinfo {author} {\bibfnamefont {Y.}~\bibnamefont {Li}}, \bibinfo
  {author} {\bibfnamefont {Y.~S.}\ \bibnamefont {You}}, \bibinfo {author}
  {\bibfnamefont {S.}~\bibnamefont {Ghimire}}, \bibinfo {author} {\bibfnamefont
  {T.~F.}\ \bibnamefont {Heinz}},\ and\ \bibinfo {author} {\bibfnamefont
  {D.~A.}\ \bibnamefont {Reis}},\ }\bibfield  {title} {\bibinfo {title}
  {High-harmonic generation from an atomically thin semiconductor},\ }\href
  {https://doi.org/10.1038/nphys3946} {\bibfield  {journal} {\bibinfo
  {journal} {Nature Physics}\ }\textbf {\bibinfo {volume} {13}},\ \bibinfo
  {pages} {262} (\bibinfo {year} {2017})}\BibitemShut {NoStop}%
\bibitem [{\citenamefont {Jensen}\ and\ \citenamefont
  {Madsen}(2022)}]{JensenPRA2022}%
  \BibitemOpen
  \bibfield  {author} {\bibinfo {author} {\bibfnamefont {S.~V.~B.}\
  \bibnamefont {Jensen}}\ and\ \bibinfo {author} {\bibfnamefont {L.~B.}\
  \bibnamefont {Madsen}},\ }\bibfield  {title} {\bibinfo {title} {Propagation
  time and nondipole contributions to intraband high-order harmonic
  generation},\ }\href {https://doi.org/10.1103/PhysRevA.105.L021101}
  {\bibfield  {journal} {\bibinfo  {journal} {Phys. Rev. A}\ }\textbf {\bibinfo
  {volume} {105}},\ \bibinfo {pages} {L021101} (\bibinfo {year}
  {2022})}\BibitemShut {NoStop}%
\bibitem [{\citenamefont {Yamada}\ and\ \citenamefont
  {Yabana}(2021)}]{Yamada2021}%
  \BibitemOpen
  \bibfield  {author} {\bibinfo {author} {\bibfnamefont {S.}~\bibnamefont
  {Yamada}}\ and\ \bibinfo {author} {\bibfnamefont {K.}~\bibnamefont
  {Yabana}},\ }\bibfield  {title} {\bibinfo {title} {Determining the optimum
  thickness for high harmonic generation from nanoscale thin films: An ab
  initio computational study},\ }\href
  {https://doi.org/10.1103/PhysRevB.103.155426} {\bibfield  {journal} {\bibinfo
   {journal} {Phys. Rev. B}\ }\textbf {\bibinfo {volume} {103}},\ \bibinfo
  {pages} {155426} (\bibinfo {year} {2021})}\BibitemShut {NoStop}%
\bibitem [{\citenamefont {Hansen}\ \emph {et~al.}(2017)\citenamefont {Hansen},
  \citenamefont {Deffge},\ and\ \citenamefont {Bauer}}]{Hansen2017}%
  \BibitemOpen
  \bibfield  {author} {\bibinfo {author} {\bibfnamefont {K.~K.}\ \bibnamefont
  {Hansen}}, \bibinfo {author} {\bibfnamefont {T.}~\bibnamefont {Deffge}},\
  and\ \bibinfo {author} {\bibfnamefont {D.}~\bibnamefont {Bauer}},\ }\bibfield
   {title} {\bibinfo {title} {High-order harmonic generation in solid slabs
  beyond the single-active-electron approximation},\ }\href
  {https://doi.org/10.1103/PhysRevA.96.053418} {\bibfield  {journal} {\bibinfo
  {journal} {Phys. Rev. A}\ }\textbf {\bibinfo {volume} {96}},\ \bibinfo
  {pages} {053418} (\bibinfo {year} {2017})}\BibitemShut {NoStop}%
\bibitem [{\citenamefont {Tancogne-Dejean}\ \emph {et~al.}(2017)\citenamefont
  {Tancogne-Dejean}, \citenamefont {M\"ucke}, \citenamefont {K\"artner},\ and\
  \citenamefont {Rubio}}]{Tancogne-Dejean2017}%
  \BibitemOpen
  \bibfield  {author} {\bibinfo {author} {\bibfnamefont {N.}~\bibnamefont
  {Tancogne-Dejean}}, \bibinfo {author} {\bibfnamefont {O.~D.}\ \bibnamefont
  {M\"ucke}}, \bibinfo {author} {\bibfnamefont {F.~X.}\ \bibnamefont
  {K\"artner}},\ and\ \bibinfo {author} {\bibfnamefont {A.}~\bibnamefont
  {Rubio}},\ }\bibfield  {title} {\bibinfo {title} {Impact of the electronic
  band structure in high-harmonic generation spectra of solids},\ }\href
  {https://doi.org/10.1103/PhysRevLett.118.087403} {\bibfield  {journal}
  {\bibinfo  {journal} {Phys. Rev. Lett.}\ }\textbf {\bibinfo {volume} {118}},\
  \bibinfo {pages} {087403} (\bibinfo {year} {2017})}\BibitemShut {NoStop}%
\bibitem [{\citenamefont {Floss}\ \emph {et~al.}(2018)\citenamefont {Floss},
  \citenamefont {Lemell}, \citenamefont {Wachter}, \citenamefont {Smejkal},
  \citenamefont {Sato}, \citenamefont {Tong}, \citenamefont {Yabana},\ and\
  \citenamefont {Burgd\"orfer}}]{Floss2018}%
  \BibitemOpen
  \bibfield  {author} {\bibinfo {author} {\bibfnamefont {I.}~\bibnamefont
  {Floss}}, \bibinfo {author} {\bibfnamefont {C.}~\bibnamefont {Lemell}},
  \bibinfo {author} {\bibfnamefont {G.}~\bibnamefont {Wachter}}, \bibinfo
  {author} {\bibfnamefont {V.}~\bibnamefont {Smejkal}}, \bibinfo {author}
  {\bibfnamefont {S.~A.}\ \bibnamefont {Sato}}, \bibinfo {author}
  {\bibfnamefont {X.-M.}\ \bibnamefont {Tong}}, \bibinfo {author}
  {\bibfnamefont {K.}~\bibnamefont {Yabana}},\ and\ \bibinfo {author}
  {\bibfnamefont {J.}~\bibnamefont {Burgd\"orfer}},\ }\bibfield  {title}
  {\bibinfo {title} {Ab initio multiscale simulation of high-order harmonic
  generation in solids},\ }\href {https://doi.org/10.1103/PhysRevA.97.011401}
  {\bibfield  {journal} {\bibinfo  {journal} {Phys. Rev. A}\ }\textbf {\bibinfo
  {volume} {97}},\ \bibinfo {pages} {011401} (\bibinfo {year}
  {2018})}\BibitemShut {NoStop}%
\bibitem [{\citenamefont {Murakami}\ \emph {et~al.}(2021)\citenamefont
  {Murakami}, \citenamefont {Takayoshi}, \citenamefont {Koga},\ and\
  \citenamefont {Werner}}]{Murakami2021}%
  \BibitemOpen
  \bibfield  {author} {\bibinfo {author} {\bibfnamefont {Y.}~\bibnamefont
  {Murakami}}, \bibinfo {author} {\bibfnamefont {S.}~\bibnamefont {Takayoshi}},
  \bibinfo {author} {\bibfnamefont {A.}~\bibnamefont {Koga}},\ and\ \bibinfo
  {author} {\bibfnamefont {P.}~\bibnamefont {Werner}},\ }\bibfield  {title}
  {\bibinfo {title} {High-harmonic generation in one-dimensional {{M}}ott
  insulators},\ }\href {https://doi.org/10.1103/PhysRevB.103.035110} {\bibfield
   {journal} {\bibinfo  {journal} {Phys. Rev. B}\ }\textbf {\bibinfo {volume}
  {103}},\ \bibinfo {pages} {035110} (\bibinfo {year} {2021})}\BibitemShut
  {NoStop}%
\bibitem [{\citenamefont {Vampa}\ \emph {et~al.}(2014)\citenamefont {Vampa},
  \citenamefont {McDonald}, \citenamefont {Orlando}, \citenamefont {Klug},
  \citenamefont {Corkum},\ and\ \citenamefont {Brabec}}]{Vampa2014}%
  \BibitemOpen
  \bibfield  {author} {\bibinfo {author} {\bibfnamefont {G.}~\bibnamefont
  {Vampa}}, \bibinfo {author} {\bibfnamefont {C.~R.}\ \bibnamefont {McDonald}},
  \bibinfo {author} {\bibfnamefont {G.}~\bibnamefont {Orlando}}, \bibinfo
  {author} {\bibfnamefont {D.~D.}\ \bibnamefont {Klug}}, \bibinfo {author}
  {\bibfnamefont {P.~B.}\ \bibnamefont {Corkum}},\ and\ \bibinfo {author}
  {\bibfnamefont {T.}~\bibnamefont {Brabec}},\ }\bibfield  {title} {\bibinfo
  {title} {Theoretical analysis of high-harmonic generation in solids},\ }\href
  {https://doi.org/10.1103/PhysRevLett.113.073901} {\bibfield  {journal}
  {\bibinfo  {journal} {Phys. Rev. Lett.}\ }\textbf {\bibinfo {volume} {113}},\
  \bibinfo {pages} {073901} (\bibinfo {year} {2014})}\BibitemShut {NoStop}%
\bibitem [{\citenamefont {Golde}\ \emph {et~al.}(2008)\citenamefont {Golde},
  \citenamefont {Meier},\ and\ \citenamefont {Koch}}]{Golde2008}%
  \BibitemOpen
  \bibfield  {author} {\bibinfo {author} {\bibfnamefont {D.}~\bibnamefont
  {Golde}}, \bibinfo {author} {\bibfnamefont {T.}~\bibnamefont {Meier}},\ and\
  \bibinfo {author} {\bibfnamefont {S.~W.}\ \bibnamefont {Koch}},\ }\bibfield
  {title} {\bibinfo {title} {High harmonics generated in semiconductor
  nanostructures by the coupled dynamics of optical inter- and intraband
  excitations},\ }\href {https://doi.org/10.1103/PhysRevB.77.075330} {\bibfield
   {journal} {\bibinfo  {journal} {Phys. Rev. B}\ }\textbf {\bibinfo {volume}
  {77}},\ \bibinfo {pages} {075330} (\bibinfo {year} {2008})}\BibitemShut
  {NoStop}%
\bibitem [{\citenamefont {Murakami}\ \emph {et~al.}(2018)\citenamefont
  {Murakami}, \citenamefont {Eckstein},\ and\ \citenamefont
  {Werner}}]{Murakami2018}%
  \BibitemOpen
  \bibfield  {author} {\bibinfo {author} {\bibfnamefont {Y.}~\bibnamefont
  {Murakami}}, \bibinfo {author} {\bibfnamefont {M.}~\bibnamefont {Eckstein}},\
  and\ \bibinfo {author} {\bibfnamefont {P.}~\bibnamefont {Werner}},\
  }\bibfield  {title} {\bibinfo {title} {High-harmonic generation in {{M}}ott
  insulators},\ }\href {https://doi.org/10.1103/PhysRevLett.121.057405}
  {\bibfield  {journal} {\bibinfo  {journal} {Phys. Rev. Lett.}\ }\textbf
  {\bibinfo {volume} {121}},\ \bibinfo {pages} {057405} (\bibinfo {year}
  {2018})}\BibitemShut {NoStop}%
\bibitem [{\citenamefont {Lysne}\ \emph {et~al.}(2020)\citenamefont {Lysne},
  \citenamefont {Murakami},\ and\ \citenamefont {Werner}}]{Lysne2020}%
  \BibitemOpen
  \bibfield  {author} {\bibinfo {author} {\bibfnamefont {M.}~\bibnamefont
  {Lysne}}, \bibinfo {author} {\bibfnamefont {Y.}~\bibnamefont {Murakami}},\
  and\ \bibinfo {author} {\bibfnamefont {P.}~\bibnamefont {Werner}},\
  }\bibfield  {title} {\bibinfo {title} {Signatures of bosonic excitations in
  high-harmonic spectra of $\text{M}$ott insulators},\ }\href
  {https://doi.org/10.1103/PhysRevB.101.195139} {\bibfield  {journal} {\bibinfo
   {journal} {Phys. Rev. B}\ }\textbf {\bibinfo {volume} {101}},\ \bibinfo
  {pages} {195139} (\bibinfo {year} {2020})}\BibitemShut {NoStop}%
\bibitem [{\citenamefont {Tancogne-Dejean}\ \emph {et~al.}(2018)\citenamefont
  {Tancogne-Dejean}, \citenamefont {Sentef},\ and\ \citenamefont
  {Rubio}}]{Tancogne-Dejean2018}%
  \BibitemOpen
  \bibfield  {author} {\bibinfo {author} {\bibfnamefont {N.}~\bibnamefont
  {Tancogne-Dejean}}, \bibinfo {author} {\bibfnamefont {M.~A.}\ \bibnamefont
  {Sentef}},\ and\ \bibinfo {author} {\bibfnamefont {A.}~\bibnamefont
  {Rubio}},\ }\bibfield  {title} {\bibinfo {title} {Ultrafast modification of
  hubbard ${{U}}$ in a strongly correlated material: Ab initio high-harmonic
  generation in {{N}}i{{O}}},\ }\href
  {https://doi.org/10.1103/PhysRevLett.121.097402} {\bibfield  {journal}
  {\bibinfo  {journal} {Phys. Rev. Lett.}\ }\textbf {\bibinfo {volume} {121}},\
  \bibinfo {pages} {097402} (\bibinfo {year} {2018})}\BibitemShut {NoStop}%
\bibitem [{\citenamefont {Imai}\ \emph {et~al.}(2020)\citenamefont {Imai},
  \citenamefont {Ono},\ and\ \citenamefont {Ishihara}}]{Imai2020}%
  \BibitemOpen
  \bibfield  {author} {\bibinfo {author} {\bibfnamefont {S.}~\bibnamefont
  {Imai}}, \bibinfo {author} {\bibfnamefont {A.}~\bibnamefont {Ono}},\ and\
  \bibinfo {author} {\bibfnamefont {S.}~\bibnamefont {Ishihara}},\ }\bibfield
  {title} {\bibinfo {title} {High harmonic generation in a correlated electron
  system},\ }\href {https://doi.org/10.1103/PhysRevLett.124.157404} {\bibfield
  {journal} {\bibinfo  {journal} {Phys. Rev. Lett.}\ }\textbf {\bibinfo
  {volume} {124}},\ \bibinfo {pages} {157404} (\bibinfo {year}
  {2020})}\BibitemShut {NoStop}%
\bibitem [{\citenamefont {Chinzei}\ and\ \citenamefont
  {Ikeda}(2020)}]{Chinzei2020}%
  \BibitemOpen
  \bibfield  {author} {\bibinfo {author} {\bibfnamefont {K.}~\bibnamefont
  {Chinzei}}\ and\ \bibinfo {author} {\bibfnamefont {T.~N.}\ \bibnamefont
  {Ikeda}},\ }\bibfield  {title} {\bibinfo {title} {Disorder effects on the
  origin of high-order harmonic generation in solids},\ }\href
  {https://doi.org/10.1103/PhysRevResearch.2.013033} {\bibfield  {journal}
  {\bibinfo  {journal} {Phys. Rev. Research}\ }\textbf {\bibinfo {volume}
  {2}},\ \bibinfo {pages} {013033} (\bibinfo {year} {2020})}\BibitemShut
  {NoStop}%
\bibitem [{\citenamefont {Orthodoxou}\ \emph {et~al.}(2021)\citenamefont
  {Orthodoxou}, \citenamefont {Za{\"i}r},\ and\ \citenamefont
  {Booth}}]{Orthodoxou2021}%
  \BibitemOpen
  \bibfield  {author} {\bibinfo {author} {\bibfnamefont {C.}~\bibnamefont
  {Orthodoxou}}, \bibinfo {author} {\bibfnamefont {A.}~\bibnamefont
  {Za{\"i}r}},\ and\ \bibinfo {author} {\bibfnamefont {G.~H.}\ \bibnamefont
  {Booth}},\ }\bibfield  {title} {\bibinfo {title} {High harmonic generation in
  two-dimensional $\text{M}$ott insulators},\ }\href
  {https://doi.org/10.1038/s41535-021-00377-8} {\bibfield  {journal} {\bibinfo
  {journal} {npj Quantum Materials}\ }\textbf {\bibinfo {volume} {6}},\
  \bibinfo {pages} {76} (\bibinfo {year} {2021})}\BibitemShut {NoStop}%
\bibitem [{\citenamefont {Shao}\ \emph {et~al.}(2022)\citenamefont {Shao},
  \citenamefont {Lu}, \citenamefont {Zhang}, \citenamefont {Yu}, \citenamefont
  {Tohyama},\ and\ \citenamefont {Lu}}]{Shao2022}%
  \BibitemOpen
  \bibfield  {author} {\bibinfo {author} {\bibfnamefont {C.}~\bibnamefont
  {Shao}}, \bibinfo {author} {\bibfnamefont {H.}~\bibnamefont {Lu}}, \bibinfo
  {author} {\bibfnamefont {X.}~\bibnamefont {Zhang}}, \bibinfo {author}
  {\bibfnamefont {C.}~\bibnamefont {Yu}}, \bibinfo {author} {\bibfnamefont
  {T.}~\bibnamefont {Tohyama}},\ and\ \bibinfo {author} {\bibfnamefont
  {R.}~\bibnamefont {Lu}},\ }\bibfield  {title} {\bibinfo {title}
  {High-harmonic generation approaching the quantum critical point of strongly
  correlated systems},\ }\href {https://doi.org/10.1103/PhysRevLett.128.047401}
  {\bibfield  {journal} {\bibinfo  {journal} {Phys. Rev. Lett.}\ }\textbf
  {\bibinfo {volume} {128}},\ \bibinfo {pages} {047401} (\bibinfo {year}
  {2022})}\BibitemShut {NoStop}%
\bibitem [{\citenamefont {Hansen}\ \emph {et~al.}(2022)\citenamefont {Hansen},
  \citenamefont {Jensen},\ and\ \citenamefont {Madsen}}]{Hansen22}%
  \BibitemOpen
  \bibfield  {author} {\bibinfo {author} {\bibfnamefont {T.}~\bibnamefont
  {Hansen}}, \bibinfo {author} {\bibfnamefont {S.~V.~B.}\ \bibnamefont
  {Jensen}},\ and\ \bibinfo {author} {\bibfnamefont {L.~B.}\ \bibnamefont
  {Madsen}},\ }\bibfield  {title} {\bibinfo {title} {Correlation effects in
  high-order harmonic generation from finite systems},\ }\href
  {https://doi.org/10.1103/PhysRevA.105.053118} {\bibfield  {journal} {\bibinfo
   {journal} {Phys. Rev. A}\ }\textbf {\bibinfo {volume} {105}},\ \bibinfo
  {pages} {053118} (\bibinfo {year} {2022})}\BibitemShut {NoStop}%
\bibitem [{\citenamefont {Masur}\ \emph {et~al.}(2022)\citenamefont {Masur},
  \citenamefont {Bondar},\ and\ \citenamefont {McCaul}}]{Masur2022}%
  \BibitemOpen
  \bibfield  {author} {\bibinfo {author} {\bibfnamefont {J.}~\bibnamefont
  {Masur}}, \bibinfo {author} {\bibfnamefont {D.~I.}\ \bibnamefont {Bondar}},\
  and\ \bibinfo {author} {\bibfnamefont {G.}~\bibnamefont {McCaul}},\
  }\bibfield  {title} {\bibinfo {title} {Optical distinguishability of
  $\text{M}$ott insulators in the time versus frequency domain},\ }\href
  {https://doi.org/10.1103/PhysRevA.106.013110} {\bibfield  {journal} {\bibinfo
   {journal} {Phys. Rev. A}\ }\textbf {\bibinfo {volume} {106}},\ \bibinfo
  {pages} {013110} (\bibinfo {year} {2022})}\BibitemShut {NoStop}%
\bibitem [{\citenamefont {Udono}\ \emph {et~al.}(2022)\citenamefont {Udono},
  \citenamefont {Sugimoto}, \citenamefont {Kaneko},\ and\ \citenamefont
  {Ohta}}]{Udono22}%
  \BibitemOpen
  \bibfield  {author} {\bibinfo {author} {\bibfnamefont {M.}~\bibnamefont
  {Udono}}, \bibinfo {author} {\bibfnamefont {K.}~\bibnamefont {Sugimoto}},
  \bibinfo {author} {\bibfnamefont {T.}~\bibnamefont {Kaneko}},\ and\ \bibinfo
  {author} {\bibfnamefont {Y.}~\bibnamefont {Ohta}},\ }\bibfield  {title}
  {\bibinfo {title} {Excitonic effects on high-harmonic generation in
  $\text{M}$ott insulators},\ }\href
  {https://doi.org/10.1103/PhysRevB.105.L241108} {\bibfield  {journal}
  {\bibinfo  {journal} {Phys. Rev. B}\ }\textbf {\bibinfo {volume} {105}},\
  \bibinfo {pages} {L241108} (\bibinfo {year} {2022})}\BibitemShut {NoStop}%
\bibitem [{\citenamefont {Murakami}\ \emph {et~al.}(2022)\citenamefont
  {Murakami}, \citenamefont {Uchida}, \citenamefont {Koga}, \citenamefont
  {Tanaka},\ and\ \citenamefont {Werner}}]{Murakami2022}%
  \BibitemOpen
  \bibfield  {author} {\bibinfo {author} {\bibfnamefont {Y.}~\bibnamefont
  {Murakami}}, \bibinfo {author} {\bibfnamefont {K.}~\bibnamefont {Uchida}},
  \bibinfo {author} {\bibfnamefont {A.}~\bibnamefont {Koga}}, \bibinfo {author}
  {\bibfnamefont {K.}~\bibnamefont {Tanaka}},\ and\ \bibinfo {author}
  {\bibfnamefont {P.}~\bibnamefont {Werner}},\ }\bibfield  {title} {\bibinfo
  {title} {Anomalous temperature dependence of high-harmonic generation in
  $\text{M}$ott insulators},\ }\href
  {https://doi.org/10.1103/PhysRevLett.129.157401} {\bibfield  {journal}
  {\bibinfo  {journal} {Phys. Rev. Lett.}\ }\textbf {\bibinfo {volume} {129}},\
  \bibinfo {pages} {157401} (\bibinfo {year} {2022})}\BibitemShut {NoStop}%
\bibitem [{\citenamefont {Hansen}\ and\ \citenamefont
  {Madsen}(2022)}]{Hansen22_2}%
  \BibitemOpen
  \bibfield  {author} {\bibinfo {author} {\bibfnamefont {T.}~\bibnamefont
  {Hansen}}\ and\ \bibinfo {author} {\bibfnamefont {L.~B.}\ \bibnamefont
  {Madsen}},\ }\bibfield  {title} {\bibinfo {title} {Doping effects in
  high-harmonic generation from correlated systems},\ }\href
  {https://doi.org/10.1103/PhysRevB.106.235142} {\bibfield  {journal} {\bibinfo
   {journal} {Phys. Rev. B}\ }\textbf {\bibinfo {volume} {106}},\ \bibinfo
  {pages} {235142} (\bibinfo {year} {2022})}\BibitemShut {NoStop}%
\bibitem [{\citenamefont {Uchida}\ \emph {et~al.}(2022)\citenamefont {Uchida},
  \citenamefont {Mattoni}, \citenamefont {Yonezawa}, \citenamefont {Nakamura},
  \citenamefont {Maeno},\ and\ \citenamefont {Tanaka}}]{Uchida2022}%
  \BibitemOpen
  \bibfield  {author} {\bibinfo {author} {\bibfnamefont {K.}~\bibnamefont
  {Uchida}}, \bibinfo {author} {\bibfnamefont {G.}~\bibnamefont {Mattoni}},
  \bibinfo {author} {\bibfnamefont {S.}~\bibnamefont {Yonezawa}}, \bibinfo
  {author} {\bibfnamefont {F.}~\bibnamefont {Nakamura}}, \bibinfo {author}
  {\bibfnamefont {Y.}~\bibnamefont {Maeno}},\ and\ \bibinfo {author}
  {\bibfnamefont {K.}~\bibnamefont {Tanaka}},\ }\bibfield  {title} {\bibinfo
  {title} {High-order harmonic generation and its unconventional scaling law in
  the $\text{M}$ott-insulating $\text{Ca}_{2}\text{RuO}_{4}$},\ }\href
  {https://doi.org/10.1103/PhysRevLett.128.127401} {\bibfield  {journal}
  {\bibinfo  {journal} {Phys. Rev. Lett.}\ }\textbf {\bibinfo {volume} {128}},\
  \bibinfo {pages} {127401} (\bibinfo {year} {2022})}\BibitemShut {NoStop}%
\bibitem [{\citenamefont {Grånäs}\ \emph {et~al.}(2020)\citenamefont
  {Grånäs}, \citenamefont {Vaskivsky}, \citenamefont {Wang}, \citenamefont
  {Thunström}, \citenamefont {Ghimire}, \citenamefont {Knut}, \citenamefont
  {Söderström}, \citenamefont {Kjellsson}, \citenamefont {Turenne},
  \citenamefont {Engel}, \citenamefont {Beye}, \citenamefont {Lu},
  \citenamefont {Reid}, \citenamefont {Schlotter}, \citenamefont {Coslovich},
  \citenamefont {Hoffmann}, \citenamefont {Kolesov}, \citenamefont
  {Schüßler-Langeheine}, \citenamefont {Styervoyedov}, \citenamefont
  {Tancogne-Dejean}, \citenamefont {Sentef}, \citenamefont {Reis},
  \citenamefont {Rubio}, \citenamefont {Parkin}, \citenamefont {Karis},
  \citenamefont {Nordgren}, \citenamefont {Rubensson}, \citenamefont
  {Eriksson},\ and\ \citenamefont {Dürr}}]{Granas2022}%
  \BibitemOpen
  \bibfield  {author} {\bibinfo {author} {\bibfnamefont {O.}~\bibnamefont
  {Grånäs}}, \bibinfo {author} {\bibfnamefont {I.}~\bibnamefont {Vaskivsky}},
  \bibinfo {author} {\bibfnamefont {X.}~\bibnamefont {Wang}}, \bibinfo {author}
  {\bibfnamefont {P.}~\bibnamefont {Thunström}}, \bibinfo {author}
  {\bibfnamefont {S.}~\bibnamefont {Ghimire}}, \bibinfo {author} {\bibfnamefont
  {R.}~\bibnamefont {Knut}}, \bibinfo {author} {\bibfnamefont {J.}~\bibnamefont
  {Söderström}}, \bibinfo {author} {\bibfnamefont {L.}~\bibnamefont
  {Kjellsson}}, \bibinfo {author} {\bibfnamefont {D.}~\bibnamefont {Turenne}},
  \bibinfo {author} {\bibfnamefont {R.~Y.}\ \bibnamefont {Engel}}, \bibinfo
  {author} {\bibfnamefont {M.}~\bibnamefont {Beye}}, \bibinfo {author}
  {\bibfnamefont {J.}~\bibnamefont {Lu}}, \bibinfo {author} {\bibfnamefont
  {A.~H.}\ \bibnamefont {Reid}}, \bibinfo {author} {\bibfnamefont
  {W.}~\bibnamefont {Schlotter}}, \bibinfo {author} {\bibfnamefont
  {G.}~\bibnamefont {Coslovich}}, \bibinfo {author} {\bibfnamefont
  {M.}~\bibnamefont {Hoffmann}}, \bibinfo {author} {\bibfnamefont
  {G.}~\bibnamefont {Kolesov}}, \bibinfo {author} {\bibfnamefont
  {C.}~\bibnamefont {Schüßler-Langeheine}}, \bibinfo {author} {\bibfnamefont
  {A.}~\bibnamefont {Styervoyedov}}, \bibinfo {author} {\bibfnamefont
  {N.}~\bibnamefont {Tancogne-Dejean}}, \bibinfo {author} {\bibfnamefont
  {M.~A.}\ \bibnamefont {Sentef}}, \bibinfo {author} {\bibfnamefont {D.~A.}\
  \bibnamefont {Reis}}, \bibinfo {author} {\bibfnamefont {A.}~\bibnamefont
  {Rubio}}, \bibinfo {author} {\bibfnamefont {S.~S.~P.}\ \bibnamefont
  {Parkin}}, \bibinfo {author} {\bibfnamefont {O.}~\bibnamefont {Karis}},
  \bibinfo {author} {\bibfnamefont {J.}~\bibnamefont {Nordgren}}, \bibinfo
  {author} {\bibfnamefont {J.~E.}\ \bibnamefont {Rubensson}}, \bibinfo {author}
  {\bibfnamefont {O.}~\bibnamefont {Eriksson}},\ and\ \bibinfo {author}
  {\bibfnamefont {H.~A.}\ \bibnamefont {Dürr}},\ }\href
  {https://doi.org/10.48550/ARXIV.2008.11115} {\bibinfo {title} {Ultrafast
  modification of the electronic structure of a correlated insulator}}
  (\bibinfo {year} {2020})\BibitemShut {NoStop}%
\bibitem [{\citenamefont {Bionta}\ \emph {et~al.}(2021)\citenamefont {Bionta},
  \citenamefont {Haddad}, \citenamefont {Leblanc}, \citenamefont {Gruson},
  \citenamefont {Lassonde}, \citenamefont {Ibrahim}, \citenamefont {Chaillou},
  \citenamefont {\'Emond}, \citenamefont {Otto}, \citenamefont
  {Jim\'enez-Gal\'an}, \citenamefont {Silva}, \citenamefont {Ivanov},
  \citenamefont {Siwick}, \citenamefont {Chaker},\ and\ \citenamefont
  {L\'egar\'e}}]{Bionta2021}%
  \BibitemOpen
  \bibfield  {author} {\bibinfo {author} {\bibfnamefont {M.~R.}\ \bibnamefont
  {Bionta}}, \bibinfo {author} {\bibfnamefont {E.}~\bibnamefont {Haddad}},
  \bibinfo {author} {\bibfnamefont {A.}~\bibnamefont {Leblanc}}, \bibinfo
  {author} {\bibfnamefont {V.}~\bibnamefont {Gruson}}, \bibinfo {author}
  {\bibfnamefont {P.}~\bibnamefont {Lassonde}}, \bibinfo {author}
  {\bibfnamefont {H.}~\bibnamefont {Ibrahim}}, \bibinfo {author} {\bibfnamefont
  {J.}~\bibnamefont {Chaillou}}, \bibinfo {author} {\bibfnamefont
  {N.}~\bibnamefont {\'Emond}}, \bibinfo {author} {\bibfnamefont {M.~R.}\
  \bibnamefont {Otto}}, \bibinfo {author} {\bibfnamefont {A.}~\bibnamefont
  {Jim\'enez-Gal\'an}}, \bibinfo {author} {\bibfnamefont {R.~E.~F.}\
  \bibnamefont {Silva}}, \bibinfo {author} {\bibfnamefont {M.}~\bibnamefont
  {Ivanov}}, \bibinfo {author} {\bibfnamefont {B.~J.}\ \bibnamefont {Siwick}},
  \bibinfo {author} {\bibfnamefont {M.}~\bibnamefont {Chaker}},\ and\ \bibinfo
  {author} {\bibfnamefont {F.}~\bibnamefont {L\'egar\'e}},\ }\bibfield  {title}
  {\bibinfo {title} {Tracking ultrafast solid-state dynamics using high
  harmonic spectroscopy},\ }\href
  {https://doi.org/10.1103/PhysRevResearch.3.023250} {\bibfield  {journal}
  {\bibinfo  {journal} {Phys. Rev. Research}\ }\textbf {\bibinfo {volume}
  {3}},\ \bibinfo {pages} {023250} (\bibinfo {year} {2021})}\BibitemShut
  {NoStop}%
\bibitem [{\citenamefont {Murakami}\ and\ \citenamefont
  {Werner}(2018)}]{Murakami2018_2}%
  \BibitemOpen
  \bibfield  {author} {\bibinfo {author} {\bibfnamefont {Y.}~\bibnamefont
  {Murakami}}\ and\ \bibinfo {author} {\bibfnamefont {P.}~\bibnamefont
  {Werner}},\ }\bibfield  {title} {\bibinfo {title} {Nonequilibrium steady
  states of electric field driven $\text{M}$ott insulators},\ }\href
  {https://doi.org/10.1103/PhysRevB.98.075102} {\bibfield  {journal} {\bibinfo
  {journal} {Phys. Rev. B}\ }\textbf {\bibinfo {volume} {98}},\ \bibinfo
  {pages} {075102} (\bibinfo {year} {2018})}\BibitemShut {NoStop}%
\bibitem [{\citenamefont {Anderson}(1958)}]{Anderson1958}%
  \BibitemOpen
  \bibfield  {author} {\bibinfo {author} {\bibfnamefont {P.~W.}\ \bibnamefont
  {Anderson}},\ }\bibfield  {title} {\bibinfo {title} {Absence of diffusion in
  certain random lattices},\ }\href {https://doi.org/10.1103/PhysRev.109.1492}
  {\bibfield  {journal} {\bibinfo  {journal} {Phys. Rev.}\ }\textbf {\bibinfo
  {volume} {109}},\ \bibinfo {pages} {1492} (\bibinfo {year}
  {1958})}\BibitemShut {NoStop}%
\bibitem [{\citenamefont {Almalki}\ \emph {et~al.}(2018)\citenamefont
  {Almalki}, \citenamefont {Parks}, \citenamefont {Bart}, \citenamefont
  {Corkum}, \citenamefont {Brabec},\ and\ \citenamefont
  {McDonald}}]{Almalki2018}%
  \BibitemOpen
  \bibfield  {author} {\bibinfo {author} {\bibfnamefont {S.}~\bibnamefont
  {Almalki}}, \bibinfo {author} {\bibfnamefont {A.~M.}\ \bibnamefont {Parks}},
  \bibinfo {author} {\bibfnamefont {G.}~\bibnamefont {Bart}}, \bibinfo {author}
  {\bibfnamefont {P.~B.}\ \bibnamefont {Corkum}}, \bibinfo {author}
  {\bibfnamefont {T.}~\bibnamefont {Brabec}},\ and\ \bibinfo {author}
  {\bibfnamefont {C.~R.}\ \bibnamefont {McDonald}},\ }\bibfield  {title}
  {\bibinfo {title} {High harmonic generation tomography of impurities in
  solids: Conceptual analysis},\ }\href
  {https://doi.org/10.1103/PhysRevB.98.144307} {\bibfield  {journal} {\bibinfo
  {journal} {Phys. Rev. B}\ }\textbf {\bibinfo {volume} {98}},\ \bibinfo
  {pages} {144307} (\bibinfo {year} {2018})}\BibitemShut {NoStop}%
\bibitem [{\citenamefont {Orlando}\ \emph {et~al.}(2018)\citenamefont
  {Orlando}, \citenamefont {Wang}, \citenamefont {Ho},\ and\ \citenamefont
  {Chu}}]{Orlando18}%
  \BibitemOpen
  \bibfield  {author} {\bibinfo {author} {\bibfnamefont {G.}~\bibnamefont
  {Orlando}}, \bibinfo {author} {\bibfnamefont {C.-M.}\ \bibnamefont {Wang}},
  \bibinfo {author} {\bibfnamefont {T.-S.}\ \bibnamefont {Ho}},\ and\ \bibinfo
  {author} {\bibfnamefont {S.-I.}\ \bibnamefont {Chu}},\ }\bibfield  {title}
  {\bibinfo {title} {High-order harmonic generation in disordered
  semiconductors},\ }\href {https://doi.org/10.1364/JOSAB.35.000680} {\bibfield
   {journal} {\bibinfo  {journal} {J. Opt. Soc. Am. B}\ }\textbf {\bibinfo
  {volume} {35}},\ \bibinfo {pages} {680} (\bibinfo {year} {2018})}\BibitemShut
  {NoStop}%
\bibitem [{\citenamefont {Yu}\ \emph {et~al.}(2019)\citenamefont {Yu},
  \citenamefont {Hansen},\ and\ \citenamefont
  {Madsen}}]{Chuan19ImperfectCrystals}%
  \BibitemOpen
  \bibfield  {author} {\bibinfo {author} {\bibfnamefont {C.}~\bibnamefont
  {Yu}}, \bibinfo {author} {\bibfnamefont {K.~K.}\ \bibnamefont {Hansen}},\
  and\ \bibinfo {author} {\bibfnamefont {L.~B.}\ \bibnamefont {Madsen}},\
  }\bibfield  {title} {\bibinfo {title} {High-order harmonic generation in
  imperfect crystals},\ }\href {https://doi.org/10.1103/PhysRevA.99.063408}
  {\bibfield  {journal} {\bibinfo  {journal} {Phys. Rev. A}\ }\textbf {\bibinfo
  {volume} {99}},\ \bibinfo {pages} {063408} (\bibinfo {year}
  {2019})}\BibitemShut {NoStop}%
\bibitem [{\citenamefont {Pattanayak}\ \emph {et~al.}(2020)\citenamefont
  {Pattanayak}, \citenamefont {S.},\ and\ \citenamefont
  {Dixit}}]{Pattanayak2020}%
  \BibitemOpen
  \bibfield  {author} {\bibinfo {author} {\bibfnamefont {A.}~\bibnamefont
  {Pattanayak}}, \bibinfo {author} {\bibfnamefont {M.~M.}\ \bibnamefont {S.}},\
  and\ \bibinfo {author} {\bibfnamefont {G.}~\bibnamefont {Dixit}},\ }\bibfield
   {title} {\bibinfo {title} {Influence of vacancy defects in solid high-order
  harmonic generation},\ }\href {https://doi.org/10.1103/PhysRevA.101.013404}
  {\bibfield  {journal} {\bibinfo  {journal} {Phys. Rev. A}\ }\textbf {\bibinfo
  {volume} {101}},\ \bibinfo {pages} {013404} (\bibinfo {year}
  {2020})}\BibitemShut {NoStop}%
\bibitem [{\citenamefont {Mrudul}\ \emph {et~al.}(2020)\citenamefont {Mrudul},
  \citenamefont {Tancogne-Dejean}, \citenamefont {Rubio},\ and\ \citenamefont
  {Dixit}}]{Mrudul2020}%
  \BibitemOpen
  \bibfield  {author} {\bibinfo {author} {\bibfnamefont {M.~S.}\ \bibnamefont
  {Mrudul}}, \bibinfo {author} {\bibfnamefont {N.}~\bibnamefont
  {Tancogne-Dejean}}, \bibinfo {author} {\bibfnamefont {A.}~\bibnamefont
  {Rubio}},\ and\ \bibinfo {author} {\bibfnamefont {G.}~\bibnamefont {Dixit}},\
  }\bibfield  {title} {\bibinfo {title} {High-harmonic generation from
  spin-polarised defects in solids},\ }\href
  {https://doi.org/10.1038/s41524-020-0275-z} {\bibfield  {journal} {\bibinfo
  {journal} {npj Computational Materials}\ }\textbf {\bibinfo {volume} {6}},\
  \bibinfo {pages} {10} (\bibinfo {year} {2020})}\BibitemShut {NoStop}%
\bibitem [{\citenamefont {Iravani}\ \emph {et~al.}(2020)\citenamefont
  {Iravani}, \citenamefont {Hansen},\ and\ \citenamefont
  {Madsen}}]{Iravani2020}%
  \BibitemOpen
  \bibfield  {author} {\bibinfo {author} {\bibfnamefont {H.}~\bibnamefont
  {Iravani}}, \bibinfo {author} {\bibfnamefont {K.~K.}\ \bibnamefont
  {Hansen}},\ and\ \bibinfo {author} {\bibfnamefont {L.~B.}\ \bibnamefont
  {Madsen}},\ }\bibfield  {title} {\bibinfo {title} {Effects of vacancies on
  high-order harmonic generation in a linear chain with band gap},\ }\href
  {https://doi.org/10.1103/PhysRevResearch.2.013204} {\bibfield  {journal}
  {\bibinfo  {journal} {Phys. Rev. Res.}\ }\textbf {\bibinfo {volume} {2}},\
  \bibinfo {pages} {013204} (\bibinfo {year} {2020})}\BibitemShut {NoStop}%
\bibitem [{\citenamefont {Pattanayak}\ \emph {et~al.}(2021)\citenamefont
  {Pattanayak}, \citenamefont {Jiménez-Galán}, \citenamefont {Ivanov},\ and\
  \citenamefont {Dixit}}]{pattanayak2021}%
  \BibitemOpen
  \bibfield  {author} {\bibinfo {author} {\bibfnamefont {A.}~\bibnamefont
  {Pattanayak}}, \bibinfo {author} {\bibfnamefont {A.}~\bibnamefont
  {Jiménez-Galán}}, \bibinfo {author} {\bibfnamefont {M.}~\bibnamefont
  {Ivanov}},\ and\ \bibinfo {author} {\bibfnamefont {G.}~\bibnamefont
  {Dixit}},\ }\bibfield  {title} {\bibinfo {title} {High harmonic spectroscopy
  of disorder-induced $\text{A}$nderson localization},\ }\href
  {https://arxiv.org/abs/2101.08536} {\bibfield  {journal} {\bibinfo  {journal}
  {arXiv}\ ,\ \bibinfo {pages} {2101.08536}} (\bibinfo {year}
  {2021})}\BibitemShut {NoStop}%
\bibitem [{\citenamefont {Zeng}\ and\ \citenamefont {Bian}(2022)}]{Zeng2022}%
  \BibitemOpen
  \bibfield  {author} {\bibinfo {author} {\bibfnamefont {A.-W.}\ \bibnamefont
  {Zeng}}\ and\ \bibinfo {author} {\bibfnamefont {X.-B.}\ \bibnamefont
  {Bian}},\ }\bibfield  {title} {\bibinfo {title} {Role of long-range
  correlations in high harmonic generation in disordered systems},\ }\href
  {https://doi.org/10.1088/1361-6455/ac5acb} {\bibfield  {journal} {\bibinfo
  {journal} {J. Phys. B}\ }\textbf {\bibinfo {volume} {55}},\ \bibinfo {pages}
  {085401} (\bibinfo {year} {2022})}\BibitemShut {NoStop}%
\bibitem [{\citenamefont {Prelov\ifmmode~\check{s}\else \v{s}\fi{}ek}\ \emph
  {et~al.}(2016)\citenamefont {Prelov\ifmmode~\check{s}\else \v{s}\fi{}ek},
  \citenamefont {Bari\ifmmode \check{s}\else \v{s}\fi{}i\ifmmode~\acute{c}\else
  \'{c}\fi{}},\ and\ \citenamefont {\ifmmode \check{Z}\else
  \v{Z}\fi{}nidari\ifmmode~\check{c}\else \v{c}\fi{}}}]{Prelov2016}%
  \BibitemOpen
  \bibfield  {author} {\bibinfo {author} {\bibfnamefont {P.}~\bibnamefont
  {Prelov\ifmmode~\check{s}\else \v{s}\fi{}ek}}, \bibinfo {author}
  {\bibfnamefont {O.~S.}\ \bibnamefont {Bari\ifmmode \check{s}\else
  \v{s}\fi{}i\ifmmode~\acute{c}\else \'{c}\fi{}}},\ and\ \bibinfo {author}
  {\bibfnamefont {M.}~\bibnamefont {\ifmmode \check{Z}\else
  \v{Z}\fi{}nidari\ifmmode~\check{c}\else \v{c}\fi{}}},\ }\bibfield  {title}
  {\bibinfo {title} {Absence of full many-body localization in the disordered
  $\text{H}$ubbard chain},\ }\href {https://doi.org/10.1103/PhysRevB.94.241104}
  {\bibfield  {journal} {\bibinfo  {journal} {Phys. Rev. B}\ }\textbf {\bibinfo
  {volume} {94}},\ \bibinfo {pages} {241104} (\bibinfo {year}
  {2016})}\BibitemShut {NoStop}%
\bibitem [{\citenamefont {Alet}\ and\ \citenamefont
  {Laflorencie}(2018)}]{Alet2018}%
  \BibitemOpen
  \bibfield  {author} {\bibinfo {author} {\bibfnamefont {F.}~\bibnamefont
  {Alet}}\ and\ \bibinfo {author} {\bibfnamefont {N.}~\bibnamefont
  {Laflorencie}},\ }\bibfield  {title} {\bibinfo {title} {Many-body
  localization: An introduction and selected topics},\ }\href
  {https://doi.org/https://doi.org/10.1016/j.crhy.2018.03.003} {\bibfield
  {journal} {\bibinfo  {journal} {Comptes Rendus Physique}\ }\textbf {\bibinfo
  {volume} {19}},\ \bibinfo {pages} {498} (\bibinfo {year} {2018})},\ \bibinfo
  {note} {quantum simulation / Simulation quantique}\BibitemShut {NoStop}%
\bibitem [{\citenamefont {Abanin}\ \emph {et~al.}(2019)\citenamefont {Abanin},
  \citenamefont {Altman}, \citenamefont {Bloch},\ and\ \citenamefont
  {Serbyn}}]{Abanin2019}%
  \BibitemOpen
  \bibfield  {author} {\bibinfo {author} {\bibfnamefont {D.~A.}\ \bibnamefont
  {Abanin}}, \bibinfo {author} {\bibfnamefont {E.}~\bibnamefont {Altman}},
  \bibinfo {author} {\bibfnamefont {I.}~\bibnamefont {Bloch}},\ and\ \bibinfo
  {author} {\bibfnamefont {M.}~\bibnamefont {Serbyn}},\ }\bibfield  {title}
  {\bibinfo {title} {Colloquium: Many-body localization, thermalization, and
  entanglement},\ }\href {https://doi.org/10.1103/RevModPhys.91.021001}
  {\bibfield  {journal} {\bibinfo  {journal} {Rev. Mod. Phys.}\ }\textbf
  {\bibinfo {volume} {91}},\ \bibinfo {pages} {021001} (\bibinfo {year}
  {2019})}\BibitemShut {NoStop}%
\bibitem [{\citenamefont {Minahan}(2006)}]{Minahan2006}%
  \BibitemOpen
  \bibfield  {author} {\bibinfo {author} {\bibfnamefont {J.~A.}\ \bibnamefont
  {Minahan}},\ }\bibfield  {title} {\bibinfo {title} {Strong coupling from the
  $\text{H}$ubbard model},\ }\href
  {https://doi.org/10.1088/0305-4470/39/41/S16} {\bibfield  {journal} {\bibinfo
   {journal} {Journal of Physics A: Mathematical and General}\ }\textbf
  {\bibinfo {volume} {39}},\ \bibinfo {pages} {13083} (\bibinfo {year}
  {2006})}\BibitemShut {NoStop}%
\bibitem [{\citenamefont {Essler}\ \emph {et~al.}(2005)\citenamefont {Essler},
  \citenamefont {Frahm}, \citenamefont {Göhmann}, \citenamefont {Klümper},\
  and\ \citenamefont {Korepin}}]{Hubbard}%
  \BibitemOpen
  \bibfield  {author} {\bibinfo {author} {\bibfnamefont {F.~H.~L.}\
  \bibnamefont {Essler}}, \bibinfo {author} {\bibfnamefont {H.}~\bibnamefont
  {Frahm}}, \bibinfo {author} {\bibfnamefont {F.}~\bibnamefont {Göhmann}},
  \bibinfo {author} {\bibfnamefont {A.}~\bibnamefont {Klümper}},\ and\
  \bibinfo {author} {\bibfnamefont {V.~E.}\ \bibnamefont {Korepin}},\ }\href
  {https://doi.org/10.1017/CBO9780511534843} {\emph {\bibinfo {title} {The
  One-Dimensional {{H}}ubbard Model}}}\ (\bibinfo  {publisher} {Cambridge
  University Press},\ \bibinfo {address} {Cambridge},\ \bibinfo {year}
  {2005})\BibitemShut {NoStop}%
\bibitem [{\citenamefont {Gaarde}\ \emph {et~al.}(2008)\citenamefont {Gaarde},
  \citenamefont {Tate},\ and\ \citenamefont {Schafer}}]{Gaarde2008}%
  \BibitemOpen
  \bibfield  {author} {\bibinfo {author} {\bibfnamefont {M.~B.}\ \bibnamefont
  {Gaarde}}, \bibinfo {author} {\bibfnamefont {J.~L.}\ \bibnamefont {Tate}},\
  and\ \bibinfo {author} {\bibfnamefont {K.~J.}\ \bibnamefont {Schafer}},\
  }\bibfield  {title} {\bibinfo {title} {Macroscopic aspects of attosecond
  pulse generation},\ }\href {https://doi.org/10.1088/0953-4075/41/13/132001}
  {\bibfield  {journal} {\bibinfo  {journal} {J. Phys. B}\ }\textbf {\bibinfo
  {volume} {41}},\ \bibinfo {pages} {132001} (\bibinfo {year}
  {2008})}\BibitemShut {NoStop}%
\bibitem [{\citenamefont {Baggesen}\ and\ \citenamefont
  {Madsen}(2011)}]{Baggesen2011}%
  \BibitemOpen
  \bibfield  {author} {\bibinfo {author} {\bibfnamefont {J.~C.}\ \bibnamefont
  {Baggesen}}\ and\ \bibinfo {author} {\bibfnamefont {L.~B.}\ \bibnamefont
  {Madsen}},\ }\bibfield  {title} {\bibinfo {title} {On the dipole, velocity
  and acceleration forms in high-order harmonic generation from a single atom
  or molecule},\ }\href {https://doi.org/10.1088/0953-4075/44/11/115601}
  {\bibfield  {journal} {\bibinfo  {journal} {J. Phys. B}\ }\textbf {\bibinfo
  {volume} {44}},\ \bibinfo {pages} {115601} (\bibinfo {year}
  {2011})}\BibitemShut {NoStop}%
\bibitem [{\citenamefont {Mahan}(2000)}]{Mahan}%
  \BibitemOpen
  \bibfield  {author} {\bibinfo {author} {\bibfnamefont {G.~D.}\ \bibnamefont
  {Mahan}},\ }\href {https://link.springer.com/book/10.1007/978-1-4757-5714-9}
  {\emph {\bibinfo {title} {Many-Particle Physics}}}\ (\bibinfo  {publisher}
  {Kluwer Academic},\ \bibinfo {address} {New York},\ \bibinfo {year} {2000})\
  p.~\bibinfo {pages} {24}\BibitemShut {NoStop}%
\bibitem [{\citenamefont {Park}\ and\ \citenamefont {Light}(1986)}]{Park1986}%
  \BibitemOpen
  \bibfield  {author} {\bibinfo {author} {\bibfnamefont {T.~J.}\ \bibnamefont
  {Park}}\ and\ \bibinfo {author} {\bibfnamefont {J.~C.}\ \bibnamefont
  {Light}},\ }\bibfield  {title} {\bibinfo {title} {Unitary quantum time
  evolution by iterative $\text{L}$anczos reduction},\ }\href
  {https://doi.org/10.1063/1.451548} {\bibfield  {journal} {\bibinfo  {journal}
  {J. Chem. Phys.}\ }\textbf {\bibinfo {volume} {85}},\ \bibinfo {pages} {5870}
  (\bibinfo {year} {1986})}\BibitemShut {NoStop}%
\bibitem [{\citenamefont {Smyth}\ \emph {et~al.}(1998)\citenamefont {Smyth},
  \citenamefont {Parker},\ and\ \citenamefont {Taylor}}]{Smyth1998}%
  \BibitemOpen
  \bibfield  {author} {\bibinfo {author} {\bibfnamefont {E.~S.}\ \bibnamefont
  {Smyth}}, \bibinfo {author} {\bibfnamefont {J.~S.}\ \bibnamefont {Parker}},\
  and\ \bibinfo {author} {\bibfnamefont {K.}~\bibnamefont {Taylor}},\
  }\bibfield  {title} {\bibinfo {title} {Numerical integration of the
  time-dependent $\text{S}$chrödinger equation for laser-driven helium},\
  }\href {https://doi.org/https://doi.org/10.1016/S0010-4655(98)00083-6}
  {\bibfield  {journal} {\bibinfo  {journal} {Comput. Phys. Commun.}\ }\textbf
  {\bibinfo {volume} {114}},\ \bibinfo {pages} {1} (\bibinfo {year}
  {1998})}\BibitemShut {NoStop}%
\bibitem [{\citenamefont {Guan}\ \emph {et~al.}(2007)\citenamefont {Guan},
  \citenamefont {Zatsarinny}, \citenamefont {Bartschat}, \citenamefont
  {Schneider}, \citenamefont {Feist},\ and\ \citenamefont {Noble}}]{Guan2007}%
  \BibitemOpen
  \bibfield  {author} {\bibinfo {author} {\bibfnamefont {X.}~\bibnamefont
  {Guan}}, \bibinfo {author} {\bibfnamefont {O.}~\bibnamefont {Zatsarinny}},
  \bibinfo {author} {\bibfnamefont {K.}~\bibnamefont {Bartschat}}, \bibinfo
  {author} {\bibfnamefont {B.~I.}\ \bibnamefont {Schneider}}, \bibinfo {author}
  {\bibfnamefont {J.}~\bibnamefont {Feist}},\ and\ \bibinfo {author}
  {\bibfnamefont {C.~J.}\ \bibnamefont {Noble}},\ }\bibfield  {title} {\bibinfo
  {title} {General approach to few-cycle intense laser interactions with
  complex atoms},\ }\href {https://doi.org/10.1103/PhysRevA.76.053411}
  {\bibfield  {journal} {\bibinfo  {journal} {Phys. Rev. A}\ }\textbf {\bibinfo
  {volume} {76}},\ \bibinfo {pages} {053411} (\bibinfo {year}
  {2007})}\BibitemShut {NoStop}%
\bibitem [{\citenamefont {Frapiccini}\ \emph {et~al.}(2014)\citenamefont
  {Frapiccini}, \citenamefont {Hamido}, \citenamefont {Schr\"oter},
  \citenamefont {Pyke}, \citenamefont {Mota-Furtado}, \citenamefont {O'Mahony},
  \citenamefont {Madro\~nero}, \citenamefont {Eiglsperger},\ and\ \citenamefont
  {Piraux}}]{Frapiccini2014}%
  \BibitemOpen
  \bibfield  {author} {\bibinfo {author} {\bibfnamefont {A.~L.}\ \bibnamefont
  {Frapiccini}}, \bibinfo {author} {\bibfnamefont {A.}~\bibnamefont {Hamido}},
  \bibinfo {author} {\bibfnamefont {S.}~\bibnamefont {Schr\"oter}}, \bibinfo
  {author} {\bibfnamefont {D.}~\bibnamefont {Pyke}}, \bibinfo {author}
  {\bibfnamefont {F.}~\bibnamefont {Mota-Furtado}}, \bibinfo {author}
  {\bibfnamefont {P.~F.}\ \bibnamefont {O'Mahony}}, \bibinfo {author}
  {\bibfnamefont {J.}~\bibnamefont {Madro\~nero}}, \bibinfo {author}
  {\bibfnamefont {J.}~\bibnamefont {Eiglsperger}},\ and\ \bibinfo {author}
  {\bibfnamefont {B.}~\bibnamefont {Piraux}},\ }\bibfield  {title} {\bibinfo
  {title} {Explicit schemes for time propagating many-body wave functions},\
  }\href {https://doi.org/10.1103/PhysRevA.89.023418} {\bibfield  {journal}
  {\bibinfo  {journal} {Phys. Rev. A}\ }\textbf {\bibinfo {volume} {89}},\
  \bibinfo {pages} {023418} (\bibinfo {year} {2014})}\BibitemShut {NoStop}%
\bibitem [{\citenamefont {Tomita}\ and\ \citenamefont
  {Nasu}(2001)}]{Tomita2001}%
  \BibitemOpen
  \bibfield  {author} {\bibinfo {author} {\bibfnamefont {N.}~\bibnamefont
  {Tomita}}\ and\ \bibinfo {author} {\bibfnamefont {K.}~\bibnamefont {Nasu}},\
  }\bibfield  {title} {\bibinfo {title} {Quantum fluctuation effects on light
  absorption spectra of the one-dimensional extended {{H}}ubbard model},\
  }\href {https://doi.org/10.1103/PhysRevB.63.085107} {\bibfield  {journal}
  {\bibinfo  {journal} {Phys. Rev. B}\ }\textbf {\bibinfo {volume} {63}},\
  \bibinfo {pages} {085107} (\bibinfo {year} {2001})}\BibitemShut {NoStop}%
\bibitem [{\citenamefont {Griffiths}(2017)}]{GriffithsQuant}%
  \BibitemOpen
  \bibfield  {author} {\bibinfo {author} {\bibfnamefont {D.~J.}\ \bibnamefont
  {Griffiths}},\ }\href@noop {} {\emph {\bibinfo {title} {Introduction to
  quantum mechanics}}},\ \bibinfo {edition} {2nd}\ ed.\ (\bibinfo  {publisher}
  {Cambridge University Press},\ \bibinfo {year} {2017})\BibitemShut {NoStop}%
\end{thebibliography}%
\end{document}